\documentclass[a4paper,11pt]{article}
\pdfoutput=1 

\usepackage{subcaption}
\usepackage{jcappub} 

\usepackage[T1]{fontenc} 
\usepackage{dsfont}
\def\be{\begin{equation}}
\def\ee{\end{equation}}
\def\bea{\begin{eqnarray}}
\def\eea{\end{eqnarray}}
\def\beq{\begin{eqnarray}}
\def\eeq{\end{eqnarray}}
\def\bas{\begin{subequations}\begin{eqnarray}}
\def\eas{\end{eqnarray}\end{subequations}}
\def\nn{\nonumber}

\def\eps{\varepsilon}

\def\la{\langle}
\def\ra{\rangle}

\def\f{\frac}

\def\SU{\text{SU}}

\def\SL{\text{SL}}

\def\sl{\mathfrak{sl}}

\def\lsim{\mathrel{\mathop  {\hbox{\lower0.5ex\hbox{$\sim$}
\kern-0.8em\lower-0.7ex\hbox{$<$}}}}}
\def\gsim{\mathrel{\mathop  {\hbox{\lower0.5ex\hbox{$\sim$}
\kern-0.8em\lower-0.7ex\hbox{$>$}}}}}
\def\cL{{\cal L}}

\def\be{\begin{equation}}
\def\ee{\end{equation}}
\def\bea{\begin{eqnarray}}
\def\eea{\end{eqnarray}}
\def\beq{\begin{eqnarray}}
\def\eeq{\end{eqnarray}}
\def\bas{\begin{subequations}\begin{eqnarray}}
\def\eas{\end{eqnarray}\end{subequations}}
\def\nn{\nonumber}

\def\eps{\varepsilon}

\def\la{\langle}
\def\ra{\rangle}

\def\f{\frac}

\def\SU{\text{SU}}

\def\SL{\text{SL}}

\def\sl{\mathfrak{sl}}

\newcommand{\N}{{\mathbb N}}
\newcommand{\R}{{\mathbb R}}
\newcommand{\Z}{{\mathbb Z}}

\newcommand{\cJ}{{\mathcal J}}
\newcommand{\cK}{{\mathcal K}}
\newcommand{\cH}{{\mathcal H}}

\newcommand{\cO}{{\mathcal O}}

\newcommand{\cV}{{\mathcal V}}
\newcommand{\cD}{{\mathcal D}}
\newcommand{\cC}{{\mathcal C}}
\newcommand{\cS}{{\mathcal S}}

\def\trho{\tilde{\rho}}

\def\pp{\partial}
\def\rd{\textrm{d}}

\def\ka{\kappa}
\def\vphi{\varphi}
\def\eps{\epsilon}

\def\la{\langle}
\def\ra{\rangle}

\newcommand{\bes}{\begin{eqnarray}}
\newcommand{\ees}{\end{eqnarray}}

\renewcommand{\sl}{{\mathfrak{sl}}}

\def\nn{\nonumber}
\def\pp{\partial}

\def\ka{\kappa}
\def\vphi{\varphi}
\def\eps{\epsilon}

\def\hv{\hat{v}}

\def\hcJ{\widehat{\cJ}}
\def\hcK{\widehat{\cK}}
\def\hcH{\widehat{\cH}}
\def\hcC{\widehat{\cC}}
\def\hcV{\widehat{\cV}}

\def\tsigma{\tilde{\sigma}}
\def\ttau{\tilde{\tau}}

\numberwithin{equation}{section}

\title{\boldmath Protected  $\SL(2,\mathbb{R})$ symmetry  \\in Quantum Cosmology }

\author[a,1]{J. Ben Achour,\note{Corresponding author.}}
\author[b,c]{E. Livine}

\affiliation[a]{Center for Relativity and Gravitation, Beijing Normal University,  Beijing 100875, China}
\affiliation[b]{Universit\'e de Lyon, ENS de Lyon,  Laboratoire de Physique, CNRS, F-69342 Lyon, France}
\affiliation[c]{Perimeter Institute, 31 Caroline Street North, Waterloo, Ontario, Canada N2L 2Y5}

\emailAdd{jbenachour@bnu.cn, etera.livine@ens-lyon.fr}

\abstract{
\\

 The polymer quantization of cosmological backgrounds provides an alternative path to the original Wheeler-de Witt (WdW) quantum cosmology, based on a different representation the commutation relations of the canonical variables. This polymer representation allows to capture the lattice like structure of the quantum geometry and leads to a radically different quantum cosmology compared to the WdW construction. This new quantization scheme has attracted considerable attention due to the singularity resolution it allows in a wide class of symmetry reduced gravitational systems, in particular where the WdW scheme fails. However, as any canonical quantization scheme, ambiguities in the construction of the quantum theory, being regularization or factor-ordering ones, can drastically modify the resulting quantum dynamics.
In this work, we propose a new criteria to restrict the quantization ambiguities in the simplest model of polymer quantum cosmology, for homogeneous and isotropic General Relativity minimally coupled to a massless scalar field. This new criteria is based on an underlying $\sl(2,\mathbb{R})$ structure present in the phase space of this simple cosmological model. By preserving the symmetry of this cosmological system under this 1d conformal group, we derive a new regularization of the phase space.
We perform both its polymer quantization and a quantization scheme directly providing a representation of the $\SL(2,\R)$ group action.
The resulting quantum cosmology can be viewed as a lattice-like quantum mechanics with an $\SL(2,\mathbb{R})$ invariance. This provides a new version of Loop Quantum Cosmology consistent with the  conformal symmetry.  This alternative construction opens new directions, among which a possible mapping with the conformal quantum mechanics as well as with recent matrix or tensor models constructions for quantum cosmological space-time.}

\begin{document}
\maketitle

\section*{Introduction}

The quantization of the Universe as a single homogeneous object is one of the simplest model of quantum gravity one can develop. By freezing a large number of degree of freedom and quantizing the single conformal mode of the metric coupled to some matter degrees of freedom, one obtains a quantum mechanical model describing the quantum evolution of an homogeneous region of space. Among the different questions to be answered in the quantum theory is the fate of the classical Big Bang singularity. It is expected that the geodesic incompleteness and thus the breakdown of the predictability at very high energy in cosmological spacetime will be replaced after quantization by a well-defined unitary quantum evolution through a fuzzy geometry. This is especially important for early cosmology predictions since a regular quantum geometry could have non-trivial effects on the initial vacuum states of cosmological perturbations, whose choice is a crucial assumption at the root of the predictions in the inflation paradigm.

However, building a consistent quantization of homogeneous region of space-time is challenging and no definitive picture has yet emerged. See \cite{Hartle:1997hw, Vilenkin:1994ua, Bojowald:2010cj, Bojowald:2015iga, Bojowald:2012xy, Bojowald:2011zzb} for different reviews. While the time-less nature of the evolution equation in quantum cosmology has triggered a intense activity to provide a consistent interpretation of the wave function of the universe (see \cite{Halliwell:1992cj, Halliwell:2002cg, Vidotto:2015bza} for interesting discussion on this issue), the different approaches to quantum cosmology have led to a variety of proposals for the initial conditions of the quantum universe. Perhaps the two most influential approaches have been the canonical Wheeler-de Witt quantization and the euclidean path integral approach which led to the famous no-boundary or tunneling proposals \cite{Hartle:1983ai, Vachaspati:1988as}. See \cite{Feldbrugge:2017kzv, Feldbrugge:2017fcc, Feldbrugge:2017mbc, Vilenkin:2018dch, Vilenkin:2018oja, Halliwell:2018ejl} for recent developments on these approaches.

The development of non-perturbative quantization technics adapted to the background independent nature of gravity, such as Loop Quantum Gravity (LQG) \cite{Thiemann:2007zz, Bodendorfer:2016uat} and the related path integral formalism as spinfoam models \cite{Reisenberger:1996pu,Perez:2012wv,Livine:2010zx}, has provided a new framework for the quantization of cosmological systems. The resulting Loop Quantum Cosmology provides a lattice-like quantization of the universe, based on a representation of the canonical commutation relation inequivalent to the standard Schrodinger representation used in the WdW scheme \cite{LQCReview, LQCReport}. This alternative \textit{polymer representation} encodes the fundamental discrete nature of the underlying geometry at high energy \cite{Ashtekar:1995mu}. Coarse-graining this quantum discrete geometry, one recovers at meso-scopic scale the WdW quantum cosmology \cite{Fredenhagen:2006wp, Corichi:2007tf}, while the classical picture is recovered in the infrared as expected. In this sense, the WdW quantization only captures the quantum physics of the universe at an intermediate range of scale between the infrared and the deep UV scales.
The lattice-like quantization obtained in LQC, where the lattice is itself subject to quantization, appears therefore as more fundamental to capture the deep UV physics. An advantage of this new non-perturbative quantization scheme is to resolve, at least for some simple cosmological models, the classical singularities and leads to regular quantum Dirac's observables in the quantum theory \cite{Bojowald:2001xe, Ashtekar:2007em, Ashtekar:2008ay, Corichi:2009pp}. Such singularity resolution mechanism does not rely on some ad hoc boundary condition on the wave function, as usually introduced in the standard the WdW scheme, but descends from the imposition of the quantum dynamics. In some specific case, this singularity resolution can leads to a bouncing dynamics\footnote{It worth pointing that such singularity resolution and the associated bounce picture is not generic in LQC. Indeed, as shown for example in \cite{Bojowald:2014qha}, the quantum fluctuations can drastically modify the bounce picture, through a backreaction of the higher moments of the wave function. In that case, the effective description commonly adopted in LQC is no longer valid and one has to perform a detailed study of the dynamics of the fluctuations to capture the dynamics beyond the effective approach. }. Moreover, the polymer representation enjoys a uniqueness theorem (similar to the Stone-Von Neumann uniqueness theorem for the standard Schrodinger representation) which ensures the robustness of the polymer quantization scheme \cite{Engle:2016zac, Ashtekar:2012cm}.  

However, as any canonical quantization, the construction of the LQC quantum theory suffers from quantization ambiguities. Additionally to the standard factor ordering issues \cite{Bojowald:2014ija, Livine:2012mh}, the regularization of the Hamiltonian constraints in term of Weyl operators inherent to the polymer quantization introduces another level of ambiguity. For different choices of regularization prior to quantization, one can obtain very different quantum physics, the best example being the closed LQC universe in which two consistent quantization schemes exist leading to drastically different dynamics \cite{Corichi:2011pg, Dupuy:2016upu, Corichi:2013usa}. See also \cite{BenAchour:2016ajk, Singh:2013ava} for the case of isotropic and anisotropic flat models. For inhomogeneous gravitational system, a standard technic to narrow to possible regularization ambiguities is to demand that the Dirac's hypersurface algebra, encoding the covariance of the system, remains anomaly-free. This has been widely used in inhomogeneous cosmology, spherical symmetry and cylindrical symmetry \cite{Barrau:2014maa, Cailleteau:2011kr, Tibrewala:2013kba, Bojowald:2015zha, Bojowald:2016vlj, BenAchour:2016brs, BenAchour:2017ivq, Bojowald:2015sta, BenAchour:2017jof, Wu:2018mhg}. However, this method cannot be used in homogeneous systems since the Dirac's algebra reduces to a trivial vanishing bracket between the scalar constraint with itself. Despite several efforts in classifying regularization (and spin) ambiguities in polymer homogeneous models \cite{Vandersloot:2005kh, Chiou:2009yx, BenAchour:2016ajk, Perez:2005fn}, it was not possible so far to find a concrete criteria to constraint the regularization ambiguities in such homogeneous mini-superspace.

The first goal of this work is precisely to introduce such new criteria. As a first step, we will only consider the homogeneous and isotropic FLRW geometry coupled to a massless scalar field. The central ingredient of our work is the existence of an underlying conformal structure in this
mini-superspace, similar to the well known conformal mechanics of de Alfaro, Fubini and Furland \cite{deAlfaro:1976vlx}. See \cite{Andrzejewski:2011ya, Andrzejewski:2015jya, Okazaki:2017lpn} for more details on this system. At the level of the phase space, this invariance under the one dimensional conformal group $\SL(2,\mathbb{R})$ fully encodes the cosmological dynamics. By demanding that this conformal structure be preserved further in the quantum theory, we obtain a new polymer regularization of the cosmological system whose regularization is constrained by conformal symmetry\footnote{The  conformal structure we consider here  is a property of homogeneous and isotropic General Relativity and is not an additional gauge invariance added to general relativity. We nevertheless point out that the loop quantization of conformal gravity has also been studied, e.g. in \cite{Campiglia:2016bhw, Wang:2018bdg}.}. 

The second goal in constructing this new $\SL(2,\mathbb{R})$-invariant realization of LQC is to discuss the compatibility of scale transformations with the existence of a minimal universal length scale at the quantum level. Indeed, quantizing this gravitational $\SL(2,\mathbb{R})$-invariant system, whatever the quantization scheme, irremediably introduces a length scale signaling the quantum gravity regime. The existence of this fundamental length scale  clashes with the  scale invariance of the system and finding a suitable quantization of the system reconciling these two aspects is a non-trivial exercise.
In standard LQC, the minimal length scale is defined by the area gap and fails to be a universal quantity, i.e.  it is not invariant under scale transformations (inverse rescaling of the induced 3d metric and the extrinsic curvature as the canonical variables for the space-time geometry). This anomaly is known as the Immirzi ambiguity, also present in the full theory of loop quantum gravity.
On the contrary, our new $\SL(2,\mathbb{R})$-invariant construction leads to a polymer quantum cosmology in which scale transformations are realized as unitary operators and the minimal length scale is invariant under  scale transformations. We further show that this new regularization turns out to be a non-linear redefinition of the classical FLRW cosmology, which maps the usual cosmological evolution with a Big Bang to a regularized cosmological evolution with a Big Bounce.
In the previous work  \cite{BenAchour:2017qpb}, we identified the $\SL(2,\R)$ symmetry in the classical FLRW cosmology and we proposed a new $\SL(2,\mathbb{R})$-invariant regularization for LQC. Here we go further and proceed to the quantization, using both a polymer quantization scheme and a $\SL(2,\R)$ group quantization scheme, and show that the resulting dilatation operator generates unitary scale transformations of quantum geometry states as wanted.

At the end of the day, the present work provides a first example of a  quantum cosmology with both a universal minimal length scale and a $\SL(2,\mathbb{R})$ conformal invariance. The conformal invariance, in such a one dimensional gravitational system, can have far reaching consequence, since it can be used to constrain the $n$-point correlations functions solely based on symmetry principles. Preserving the $\SL(2,\mathbb{R})$ structure could therefore open the possibility to bootstrap this quantum cosmology, leaving aside the knowledge of the quantum corrections at the Hamiltonian or Lagrangian level, to fully solve the quantum theory based on conformal invariance arguments. This provides another motivation for preserving the $\SL(2,\mathbb{R})$ structure through the quantization process.

\medskip

The layout of this article is as follow. In section-\ref{A}, we review the symmetry of the homogeneous and isotropic Einstein-Scalar system under the one dimensional conformal group $\SL(2,\mathbb{R})$, at the level of the phase space. 
In section-\ref{B1}, we first show how the standard LQC regularization breaks this conformal structure and we derive an improved $\SL(2,\mathbb{R})$-preserving regularization which leads to a new regularized Hamiltonian (\ref{Hilqc}). In particular,  Section-\ref{B3} is devoted to the derivation of the modified Friedman equations and to the resulting effective bouncing dynamics avoiding the initial cosmological singularity.
In Section-\ref{B2}, we present the loop quantization of the new regularized cosmological system and discuss the deparametrized dynamics and the associated singularity resolution.
Finally, in Section-\ref{C}, we proceed to a group quantization preserving the full $\SL(2,\R)$ group action on the cosmological phase space. We  compare this alternative quantization scheme to this polymer quantization and discuss its advantage to quantize the inverse volume term encoding the coupling between the gravitational and matter degrees of freedom.
We conclude by a discussion of potential generalization of our results in Section-\ref{D}. The $\SL(2,\mathbb{R})$ realization of polymer quantum cosmology developed here is briefly summarized in a short companion letter \cite{BenAchour:2018jwq}. 

Additionally, we show in appendix that the  conformal symmetry also applies to the Bianchi I  model for homogeneous cosmology, hinting that the $\SL(2,\R)$ symmetry is more general than the homogeneous and isotropic context of the FLRW cosmology and that the general conditions for its existence should be studied in more details in general relativity.

\section{$\SL(2,\mathbb{R})$ structure in FLRW cosmology}
\label{A}

We begin be reviewing briefly the conformal structure encoded in the homogeneous and isotropic sector of the Einstein-Scalar system,  presented in \cite{BenAchour:2017qpb}. 

\subsection{Scaling properties of the space-like hypersurface}

Let us start with a homogeneous massless scalar field minimally coupled to an FLRW  background with flat slices. The metric is given by 
\be 
ds^2 = - N^2(t)\rd t^2 + a^2(t) \delta_{ij} \rd x^i \rd x^j
\ee 
We integrate the Einstein-Hilbert action for gravity coupled to a homogeneous massless free scalar field over a fiducial 3D-cell of volume $V_{\circ}$,  which gives the reduced  action for the Einstein-Scalar system:
\begin{align}
\label{actionFRW}
\cS[a, N, \phi] & = V_{\circ} \int dt \left[ - \frac{3}{8\pi G} \frac{a \dot{a}^2}{N} + \frac{a^3}{2N} \dot{\phi}^2 \right]
\,.
\end{align}
We compute the conjugate momenta to the variables $(a, \phi)$,
\be
\pi_a = \frac{\delta \cL}{\delta \dot{a}} 
= -\frac{3 V_{\circ}}{4\pi G} \frac{a \dot{a} }{N}
\,,\qquad
\pi_{\phi} = \frac{\delta \cL}{\delta \dot{\phi}} =  \frac{a^3 V_{\circ}}{N} \dot{\phi}
\,,
\ee
so the reduced action describes a constrained system, $\cS=\int dt\,[\dot{a}\pi_{a}+\dot{\phi}\pi_{\phi}-H[N]]$ with the scalar constraint given by\footnotemark:
\be
H [N] =  \f N {V_{\circ}} \left( \frac{\pi^2_{\phi}}{2 a^3} - \frac{2\pi G}{3} \frac{\pi^2_a}{a}\right)
\,.
\ee
\footnotetext{
One can absorb the factor $V_{\circ}$ in the definition of the lapse $N$ without affecting the physics. We should nevertheless point out that the Hamiltonian $H$ has dimension $[L]^{-1}$ since the lapse $N$ is dimensionless. If we absorb the fiducial volume in the lapse, the Hamiltonian then seems to change dimension to $[L]^2$ if we forget that the lapse has acquired a physical dimension.}
It is useful to work instead with a different set of canonical variables which are the 3d volume $v$ of the fiducial cell
and the extrinsic curvature (or  Hubble rate) $b$, with Poisson bracket $\{ b, v\} = 1$. They are related to the scale factor $a$ and its conjugate momentum $\pi_{a}$ by 
\be
v = a^3\,V_{\circ}
\,,
 \qquad 
 b = - \frac{1}{V_{\circ}} \frac{ \pi_a}{a^2} =  \frac{1}{4\pi G}\frac{\dot{a}}{Na}
 \,.
\ee
We work in units where we have set the Planck constant $\hbar$ and the speed of light $c$ to 1. So $G$ has the dimension of a squared length and actually defines the Planck length $\ell_{Planck}\propto \sqrt{G}$. Hence, the canonical rescaled variables have dimension $[v] =L^{3}$ and $[b] = L^{-3}$.

The matter content is encoded in the scalar field and its conjugated momentum, with the canonical Poisson bracket $\{ \phi, \pi_{\phi} \} = 1$.
The scalar field physical dimension is an inverse length, $[\phi]=L^{-1}$, while its momentum goes as a length,  $[\pi_{\phi}]=L$.
The Hamiltonian constraint, generating the invariance under time reparametrization of the coupled geometry-matter system, is given by:
\be
\label{Hclass}
H [N]= H_{\text{m}} [N] + H_{\text{g}} [N]
=
\f N2 \left(\frac{\pi^2_{\phi}}{ v} - \ka^2 v b^2\right)
\,,
\quad\textrm{with}\quad
\ka=\sqrt{12\pi G}
\,,
\ee
where we check that both terms have the physical dimension and $H$ goes as expected as an inverse length, $[H]=L^{-1}$.
Consider now the integrated trace of the extrinsic curvature
\be
\int_{V_{\circ}} d^3x \, \sqrt{|\gamma|} \gamma^{\mu\nu} K_{\mu\nu}
=
\int  d^3x \, a^3\f{3\dot{a}}{Na}
=
\ka^2 C
\quad\textrm{with}\quad
C=bv
\,,
\ee 
where the integration has been performed over the fiducial 3D cell and $\gamma_{\mu\nu}$ is the induced metric on the space-like hypersurface $\Sigma$. 
We recognize this integrated extrinsic curvature observable $C$ as the generator of dilatations on the phase space $(b,v)$:
\be
e^{\{ \eta C, \cdot\}} \triangleright b = e^{-\eta} b
\,,\qquad
e^{\{ \eta C, \cdot\}} \triangleright v = e^{\eta} v
\,.
\ee 
It generates constant rescaling of the hypersurface. This generator is also called the complexifier in the standard LQG jargon, this name being motivated by its initial use to construct a Wick transform between the real and the self dual version of LQG and deal with the old problem of the reality condition \cite{Thiemann:1995ug, MenaMarugan:1997xj}. It has also played a crucial role in constructing coherent states in LQG \cite{Thiemann:2000bw}. In this work, we shall interchangeably use the name dilatation generator or complexifier when refering to the generator $\cC$. Now, it turns out that this generator gives the Hamiltonian evolution of the volume,
\be
\label{dil}
\frac{\rd v}{\rd \tau} = \{ v, H[N]\} = \ka^2\,C[N]
\ee
in terms of the proper time $\rd\tau = N\,\rd t$. 
Additionally, the two other brackets involving $\left( C, v , H\right)$ form a closed algebra
\be
\label{23}
\frac{\rd C}{\rd \tau} = \{ C, H[N] \} = - H[N] \qquad \{ C, v \} = v
\ee
The first equality reflects that the matter and gravitational Hamiltonians transform homogeneously under scale transformations, a special property of this simple cosmological model,
\be
\label{weightH}
e^{\{ \eta C, \cdot\}} \triangleright H = e^{-\eta} H
\,.
\ee
On shell,  scale transformations turn out to be a global symmetry of the system. This CVH algebra encodes how the volume and the Hamiltonian change with the scale and thus reflects the whole cosmological classical dynamics. Moreover, the deparametrized Hamiltonian with respect to  the clock $\phi$  coincides on-shell with the dilatation generator. Indeed, solving the Hamiltonian constraint \eqref{Hclass} $H=0$ gives
\be
\pi_{\phi} = \epsilon \ka \, C
\ee
where the sign $\epsilon = \pm 1$ corresponds to the two expanding and contracting branches of the cosmological evolution and the constant $\ka$ is simply the Planck length up to the numerical factor $\sqrt{12 \pi}$.
Therefore, the deparametrized dynamics can be identified with the dilatation transformations generated by the trace of extrinsic curvature. Finally, consider the phase space functions
\be
\cO_b = b e^{\epsilon \ka\phi} \quad \text{and} \quad \cO_v = v e^{-\epsilon \ka\phi} \,,
\qquad\textrm{with}\qquad
\cO_{b}\cO_{v}=bv=C
\,.
\ee
They are two weakly commuting Dirac's observables\footnotemark{} once one picks up a given branch corresponding to the contracting or expanding phases, i.e given a choice of sign $\epsilon$. Notice that they form a canonically conjugated pair of variables on the reduced phase space, such that $\{ \cO_b, \cO_v\}= 1$.
\footnotetext{
However, in order to find strongly commuting Dirac's observables for this cosmological system, one can combine the two expanding and contracting branches  and work with
\begin{align}
\cD_1 = \frac{1}{2\kappa} \left[ \left( \pi_{\phi} + \kappa vb\right) e^{\kappa \phi} + \left( \pi_{\phi} - \kappa vb\right) e^{-\kappa \phi} \right] \\
\cD_2 = \frac{1}{2\kappa} \left[ \left( \pi_{\phi} + \kappa vb\right) e^{\kappa \phi} - \left( \pi_{\phi} - \kappa vb\right) e^{-\kappa \phi} \right] 
\nn
\end{align}
which have exactly  vanishing Poisson brackets with the Hamiltonian constraints:
\begin{align}
\{ H, \cD_1\} = \{ H, \cD_2\} = \{ \cD_1, \cD_2\} = 0
\nn
\end{align}
}

\subsection{The  $\sl(2,\mathbb{R})$  structure}

The CVH algebra is isomorphic to the $\sl(2,\R)$ Lie algebra. Indeed, we can rearrange the generators of the CVH algebra to map it to the standard basis of the $\sl(2,\R)$ algebra:
\be
\label{su11}
k_y=C
\qquad
 k_x =\f1\ka\bigg{[}
 \frac{v}{2\sigma \ka^2} +  {\sigma \ka^2}\, H 
\bigg{]}
\qquad
j_z =
\f1\ka\bigg{[}
\frac{v}{2\sigma \ka^2} -  {\sigma \ka^2}\, H 
\bigg{]}
\,.
\ee
The squared Planck length factor $\ka^2$ is here to ensure that the volume term $\ka^{-3}v$ and the Hamiltonian constraint term $\ka H$ are both dimensionless. Then the factor $\sigma\in\R$ is an arbitrary dimensionless constant, which does not affect the $\sl(2,\R)$ symmetry. This arbitrary coupling constant also appears in conformal quantum mechanics \cite{deAlfaro:1976vlx}.
Rescaling $\sigma$  corresponds to acting with a dilatation generated by the complexifier $C=k_{y}$. It can be interpreted as a choice of length unit for the volume and the Hamiltonian, or as a choice of origin $\phi_{0}$ for the scalar field clock $\phi$ as we will comment later in section \ref{B3}.
The limit $\sigma\ka^2\rightarrow 0$ can be interpreted as the vanishing Planck length limit $\ka \rightarrow 0$ corresponding to a classical or no-gravity regime.

Then, one recovers the standard $\sl_{2}$ commutations relations from the canonical bracket $\{ b,v\}=1$,
\begin{align}
\label{CR}
\{ j_z, k_x\}  = k_y
\,, \qquad
\{ j_z, k_y\}  = - k_x
\,, \qquad
\{ k_x, k_y\}  = - j_z 
\,.
\end{align}
The $\sl_{2}$ Casimir is then given by the matter content of the universe:
\be
\label{CazMatt}
\mathfrak{C}_{\mathfrak{sl}_{2}}
= j^2_z - k^2_x - k^2_y   = -\f2{\ka^2}vH-C^{2}
= -  \frac{\pi^2_{\phi}}{\ka^2} < 0
\ee
At the classical level, this $\sl(2,\R)$ structure means that the cosmological trajectories can be integrated as $\SL(2,\R)$ transformations generated by $H$ \cite{BenAchour:2017qpb}.
At the quantum level, the Casimir formula means that the system is described by space-like representations of $\SL(2,\R)$ (i.e. from the continuous series). 

The above CVH structure actually encodes an underlying conformal symmetry of the homogeneous and isotropic action of the Einstein-Scalar system (without self interaction). This new symmetry can be understood as a type of Mobius covariance of the cosmological action (\ref{actionFRW}) very similar to the well known Mobius symmetry of conformal mechanics \cite{deAlfaro:1976vlx}. Hence, the simplest cosmological sector is invariant under the one dimensional conformal group, i.e $\SL(2,\mathbb{R})$ (or equivalently $\SU(1,1)$). This symmetry fully determines the cosmological dynamics. The details on this new symmetry of the homogeneous cosmological action will be presented in a future work \cite{CFT}. 

In the present work, our goal is to use this symmetry to build an LQC model where the generator of the scale transformation is implemented as a unitary operator at the quantum level, a property which is not present in standard LQC. This will ensure that the regularization (or polymer) scale commonly introduced in LQC remains a universal scale. Concretely, this requires to preserve the CVH algebra at the quantum level when performing the polymer quantization. This is why we focus solely on the hamiltonian CVH structure in the present work.

Nevertheless, let us emphasize that this $\sl(2,\mathbb{R})$ structure is a property of the flat homogeneous and isotropic symmetry reduced action of the Einstein-Scalar system. As such, any quantization schemes applied to this cosmological system aiming at obtaining a unitary operator of scale transformation should take into account this hidden symmetry. 
This is true for the standard Wheeler-de Witt quantization based on the Schrondinger representation, but also for the loop quantization based on the polymer representation. Moreover, in both case, this new symmetry provides a useful criteria to constrain the factor ordering and regularization ambiguities inherent to these quantizations based on the canonical procedure.

Finally, we conclude this section by pointing the fact that this symmetry is actually not a coincidence of the high degree of symmetry of the flat homogeneous and isotropic sector considered here. On the contrary, this symmetry holds for a large class of flat cosmology and massless matter coupling. In Appendix-\ref{BianchiI}, we show this $\sl(2,\mathbb{R})$ structure is still valid for the Bianchi I cosmology filled with a massless scalar field. This suggests that the improved loop regularization and the subsequent polymer quantization that we present in the next section can be easily generalized to the Bianchi I model, including thus anisotropies. 

\section{Improved loop regularization}
\label{B1}

In this section, we show as a first step that the standard regularization procedure used in LQC fails to preserve the conformal symmetry of FRW cosmology and breaks  the $\SL(2,\mathbb{R})$ symmetry induced by the CVH algebra. As a second step, we derive a minimal extension of the standard regularization  allowing to preserve the CVH algebra structure, thus making 3d scale transformations compatible with the introduction of the universal length scale $\lambda$.

\subsection{Standard regularization}

LQC consists in the polymer quantization of the homogeneous and isotropic sector of GR written in term of the Ashtekar-Barbero triad-connection variables \cite{LQCReview}.
One can either starts from the Ashtekar's phase space of homogeneous and isotropic geometries and regularize the curvature of the Ashtekar-Barbero connection by its expression in term of the holonomy of the connection around tiny, but not infinitesimal, loops. Then, one has to proceed to a suitable change of variables to arrive at the regularized phase space in term of the variables $(b,v)$. One can then proceed to the quantization in the polymer representation. This is the standard procedure adopted in LQC. See \cite{LQCReport} for a review. Equivalently, one can starts directly from the classical $(b,v)$ phase space and proceed just as the polymer quantization of the harmonic oscillator or the free particle \cite{Corichi:2007tf}.

In any case, this quantization requires a regularization of the scalar constraint. For the simple system of a massless scalar field minimally coupled to flat FLRW geometry, one can either introduce the polymer regularization at the classical level or derive it from the effective dynamics of coherent quantum states \cite{Bojowald:2006gr}. In both case, one obtains the regularized LQC Hamiltonian constraint:
\be
\label{HLQC}
\cH^{(\lambda)}_{\text{LQC}} [N] = 
\f N2\,\left(\frac{\pi^2_{\phi}}{v} - \ka^2  v \frac{\sin^2{( \lambda b)}}{\lambda^2}\right)
\,,
\ee
where the volume regularization scale,  $[\lambda ] = L^{3}$, is at the Planck scale,
\be
\lambda = \Delta \ell_{\text{Planck}}^3\,.
\ee 
The factor $\Delta$ is a proportionality constant which is not determined by first principles in loop quantum cosmology\footnote{In the standard treatment, $\Delta$ is usually fixed by demanding that it matches the volume (or area) gap derived in kinematical LQG, introducing thus an ad hoc dependency on the Immirzi parameter. Hence, the fact that the observables in this simple LQC model depend in the end on the Immirzi parameter is artificial and without this choice for $\Delta$, the model could be quantized without any reference to the Immirzi parameter.}.  
This non-perturbative quantum correction arising in the polymer quantization effectively introduces a cut-off in the extrinsic curvature, 
\be
b \leqslant \frac{\pi}{\lambda}
\ee 
which can be extremely large since $\lambda$ is a priori at the Planck scale 
Notice that the standard polymer regularization affects only the gravitational part since only the variable $b$ is polymerized.

When the Hubble factor $b$ becomes sufficiently small, which corresponds to low energy matter density in the universe and thus to the regime of large volume, we recover the standard FLRW Hamiltonian:
\be
v \frac{\sin^2{( \lambda b)}}{\lambda^2}
\,\underset{b \ll \lambda^{-1}}\longrightarrow\,
vb^2
\,,
\quad
\cH_{\text{LQC}}^{(\lambda)}[N]
\,\underset{b \ll \lambda^{-1}}\longrightarrow\,
H[N]
\,.
\ee
In the standard LQC framework, the dilatation generator is considered unchanged, i.e. we keep on working with $\cC= vb$. Under its exponentiated action, the scalar Hamiltonian constraint is not homogeneous and scale transformations actually change the regularization scale $\lambda$. More precisely, the scalar constraint $\cH^{(\lambda)}$ does not transform as earlier in \eqref{weightH}, but as
\be
e^{\{ \eta \;  \cC, \cdot\}} \triangleright \cH_{\text{LQC}}^{(\lambda)} [v,b]= \cH_{\text{LQC}}^{({\lambda'})} [v',b']
\ee
with
\begin{align}
{\lambda}'  = e^{- \eta} \lambda
\,,\qquad
v'  = e^{\eta} v
\,,\qquad
b' & = e^{- \eta} b
\,.
\end{align}
Hence, the dilatations are not consistent with the regularization scheme: they act on the scale $\lambda$ and spoil the three dimensional scale invariance of the classical cosmological model. Hence $\lambda$ is not a universal area invariant under these transformations. As a consequence, at the quantum level, the Hilbert space of quantum states at fixed $\lambda$ can not carry a unitary representation of those (3d) scale transformations. This is the origin of the Immirzi ambiguity in Loop Quantum Gravity \cite{Rovelli:1997na}.

Instead of working with the classical dilation generator, it seems more natural to preserve the symmetry of the model and adapt the CVH algebra by extending the regularization scheme to the volume, the dilatation generator and other cosmological observables. Looking back to \eqref{dil}, the dilatation generator can be obtained as the Poisson bracket of the volume and the Hamiltonian constraint, as already noticed in \cite{Thiemann:1996aw}. A natural prescription is thus:
\be
\label{RCompl}
\cC_{\text{LQC}}^{(\lambda)} = \{ v, \cH_{\text{LQC}}^{(\lambda)}\} = v \frac{\sin{\left( 2 \lambda b\right)}}{2 \lambda}
\,,
\ee
which gives back the standard dilatation generator $C=vb$ when $b \ll \lambda^{-1}$. As expected, the extrinsic curvature $b$ becomes bounded. However this still breaks the CVH algebra,
\be
\{ \cC_{\text{LQC}}^{(\lambda)}, v\} \neq v
\,,
\qquad
\{ \cC_{\text{LQC}}^{(\lambda)}, \cH_{\text{LQC}}^{(\lambda)} \} \neq - \cH_{\text{LQC}}^{(\lambda)}
\,,
\ee
and does not restore the scale invariance of FLRW cosmology. In particular, the matter and gravitational parts of the Hamiltonian do not scale in the same way. What's missing is a suitable regularization of both the volume $v$ and the inverse volume factor entering the matter Hamiltonian. 

\subsection{New $\sl(2,\mathbb{R})$-preserving regularization}

A systematic approach is to allow for a general regularization, such that the CVH generators can be written as
\begin{align}
\cH[N] = \f N2 \left(\frac{\pi^2_{\phi}}{ v f_2(b)} -\ka^2 v f_1(b)\right)
\,,\qquad
 \cC = v f_3(b)\,, \qquad \cV = v f_4(b)\,.
\end{align}
where $f_1$, $f_2$, $f_3$, $f_4$ are unknown $b$-dependent regularization functions involving implicitly the scale $\lambda$. By computing the CVH Poisson brackets and requiring that they form a closed $\sl_{2}$ algebra, we obtain a set of anomaly-free conditions:
\begin{align}
 \frac{f_3}{f_1} \pp_{b}f_1 - \pp_{b}f_3  = 1
 \,,\qquad
\pp_{b}f_3 - \frac{f_3 }{f_2} \pp_{b}f_2  = 1
\,,\qquad
\pp_{b}f_3 -  \frac{f_3 }{f_4} \pp_{b}f_4  = 1
\,,
\end{align}
while the last brackets impose
\be
f_4 \pp_{b}f_2 = f_2 \pp_{b}f_4
\,,
\qquad
f_4 \pp_{b}f_1 - f_1 \pp_{b}f_4 = 2 f_3
\,.
\ee
Choosing $f_1$ to be the standard sine square holonomy correction, i.e. 
\be 
\label{f1}
f_1(b) = \frac{\sin^2{(\lambda b)}}{\lambda^2}
\ee
as above in \eqref{HLQC}, and taking $f_2= f_4$, one obtains a solution for $f_3$, from which one can solve for $f_2$ and thus $f_4$. Checking the consistency with the last equation, one obtains that
\be
f_3(b) = \frac{\sin{(2\lambda b)}}{2 \lambda}\,, \qquad f_2(b) = f_4(b) = \cos^2(\lambda b)
\,.
\ee 
This provides the minimal polymer regularization consistent with the algebra:
\be
\label{regCVH1}
\cC = v \frac{\sin{(2 \lambda b)}}{2\lambda}\,, \qquad \cV = v \cos^2{(\lambda b)}
\ee
while the new Hamiltonian constraint reads:
\be
\label{Hilqc}
\cH_{(\textrm{new})}[N]
=\f N2 \left(
\frac{\pi^2_{\phi}}{ v \cos^2{(\lambda b)}} - \ka^2 v \frac{\sin^2{\left( \lambda b\right)}}{\lambda^2}
\right)
\,.
\ee
This gives the new effective Hamiltonian, which we introduced in \cite{BenAchour:2017qpb}. In the rest of the paper, we drop the subscript $(\textrm{new})$ and we focus on studying the classical and quantum dynamics induced by this new regularized cosmological Hamiltonian.

By construction, these regularized observables, $\cV$, $\cC$ and $\cH$, form a closed $\sl(2,\R)$ algebra:
\be
\label{LQC-CVH}
\{\cC,\cH\}=-\cH
\,,\quad
\{\cC,\cV\}=+\cV
\,,\quad
\{\cV,\cH\}=\ka^2\,\cC
\,.
\ee
We can write these in terms of the usual $\sl(2,\R)$ basis by introducing the dimensionful rescaling factor $\sigma\ka^2$:
\be
\label{LQC-sl2R}
\cC = \cK_y
= \frac{v\sin{(2 \lambda b)}}{2 \lambda} 
\,,\qquad
 \cV = {\sigma\ka^3}( \cK_x + \cJ_z)
\,,\qquad
\cH = \frac{1}{2\sigma\ka} \big{(}\cK_x - \cJ_z\big{)}\,,
\ee
\begin{align}
\cJ_z 
&=
\f v{2\sigma\ka^3}\left(
\cos^2\lambda b +\sigma^2\f{\ka^6}{ \lambda^2}\sin^2\lambda b
\right)
-
\sigma\ka\frac{ \pi^2_{\phi}}{2 v \cos^2{(\lambda b)}}
\,,\\
\cK_{x}
&=
\f v{2\sigma\ka^3}\left(
\cos^2\lambda b -\sigma^2\f{\ka^6}{ \lambda^2}\sin^2\lambda b
\right)
+
\sigma\ka\frac{ \pi^2_{\phi}}{2 v \cos^2{(\lambda b)}}
\,.
\end{align}
Notice that these loop regularized generators reproduce the Casimir equation (\ref{CazMatt}) such that it is again given by minus the rescaled kinetic energy of the scalar field. Moreover, if the regularization scale $\lambda$ and the length factor $\sigma \ka$ defining the $\sl_{2}$ structure from the CVH observable algebra are chosen to coincide, $\lambda=\sigma \ka^3$, then we recover the expressions introduced in \cite{BenAchour:2017qpb}:
\begin{align}
\label{jz}
\cJ_z  =
\frac{v}{2 \lambda} 
-
\frac{\lambda  \pi^2_{\phi}}{2\ka^2 v \cos^2{(\lambda b)}}
\,,\qquad
\cK_x  = 
\frac{v\cos{(2 \lambda b)}}{2 \lambda}
+
\frac{\lambda \pi^2_{\phi}}{2\ka^2 v \cos^2{(\lambda b)}}
\,.
\end{align}
This CVH algebra encapsulates how the theory transforms under $\sl(2,\mathbb{R})$ transformations of the time coordinate. Now the modified dilatation generator $\cC$ simply rescales the regularized Hamiltonian constraint while keeping the area scale $\lambda$ fixed:
\be
e^{\{ \eta\, \cC,\, \cdot\}} \triangleright \cH^{(\lambda)} [v,b]= e^{- \eta}\,\cH^{(\lambda)} [v,b]
\,.
\ee
Moreover the regularized volume $\cV$, which also enters the inverse volume factor of the matter Hamiltonian, has acquired a holonomy correction and simply rescales under modified dilatations. This exact scale invariance with a regularized dilatation generator (or complexifier in the LQG jargon) defined at fixed  $\lambda$ is a new feature of LQC. This improved version of LQC allows to consider $\lambda$ as a true universal fundamental scale, invariant under global scale transformations. We point that all the ambiguities have not been removed by the conformal structure since through the choice of correction $f_1$ in (\ref{f1}), the choice of spin to compute the holonomy correction remains. However, once this correction is fixed, the other regularization descend from the conformal structure. Finally, we point that the ambiguity in the choice of regularization scheme, being the curvature or connection regularization, does not affect the dynamics since both regularization schemes can be made to coincide for this flat FLRW system. See \cite{BenAchour:2016ajk} for details.

Let us now compare this new extended regularization scheme to the standard polymer regularization of the effective LQC framework.
First, in the  limit $ b \ll \lambda^{-1}$ which corresponds to small energy density for the universe, all the observables, $\cH$, $\cV$ and $\cC$,  coincide with their classical expressions, $H$, $v$ and $C$.
Then we notice that the new Hamiltonian $\cH$ is actually equivalent to the usual LQC Hamiltonian $\cH_{\textrm{LQC}}$ up to a redefinition of the lapse:
\be
\cH^{(\lambda)}[N]
=
\cH_{\textrm{LQC}}^{(2\lambda)}\big{[}N\cos^{-2}\lambda b\big{]}
\,.
\ee
This should correspond to a different definition of the lapse in terms of the 3+1 decomposition of the 4d metric. Let us now show that the above $\sl(2,\mathbb{R})$-preserving regularization defines actually a non-linear redefinition of classical FLRW cosmology.

\subsection{Non-linear redefinition of FRLW cosmology}
\label{nonlinear}

As we have introduced a regularized volume observable $\cV$, we can further define its conjugate variable:
\be
\label{newvar}
B = \frac{\tan{\left( \lambda b\right)}}{\lambda}
\,,\qquad
\cV = v \cos^2{(\lambda b)}
\,,\qquad
\{ B, \cV\} = 1
\,.
\ee
It turns that this canonical transformation from the classical to the effective cosmological phase space maps the classical FLRW model to the new Hamiltonian scalar constraint \eqref{Hilqc}:
\be
\label{nonlinearFRW}
\cH
=
\frac{\pi^2_{\phi}}{2 \cV} - \frac{\ka^2}{2} \cV B^2
\ee
However, in term of the $(B, \cV)$ canonical variables, the new effective scalar constraint takes the same expression as the classical one in term of the variables $(b,v)$.

This new pair of canonical variables $(B,\cV)$ transform as the usual ones $(b, v)$ in classical FLRW cosmology under dilatations, now generated by the regularized complexifier,
\be
\cC= \frac{v\sin{(2 \lambda b)}}{2 \lambda} =\cV B\,,\quad
e^{\{ \eta \,  \cC, \cdot\}} \triangleright B = e^{-\eta} B
\,,\quad
e^{\{ \eta \, \cC,\cdot\}} \triangleright \cV = e^{\eta} \cV
\,.
\ee
This shows that this new version of LQC is simply a non-linear redefinition of FLRW cosmology, preserving its scale invariance and $\SL(2,\R)$ symmetry. This parallels the construction of doubly special relativity with a non-linear action of the Lorentz group compatible with the fundamental Planck length, thereby deforming but not breaking the Lorentz symmetry of flat space-time \cite{AmelinoCamelia:2002wr}.

This means that the regularized volume $\cV$ will follow the usual FLRW cosmological trajectories.
We can go further and give the on-shell relation between the 3d volume $v$ and the regularized volume $\cV$. Indeed, since $\cC$ is a constant of motion, further equal to $\pi_{\phi}/\ka$ as soon as the Hamiltonian vanishes $\cH=0$, we have the following identity:
\be
\label{vVV}
{v\cos \lambda b\sin\lambda b}=\lambda\cC
\quad\Rightarrow\quad
v=v\cos^2 \lambda b+v\sin^2\lambda b
=
\cV+\f{\lambda^2\cC^2}\cV\,.
\ee
Although the regularized volume $\cV$ will follow the usual trajectories of the classical FLRW cosmology, starting from the singularity at $\cV=0$ and growing to infinity $\cV\rightarrow+\infty$, this non-linear relation means that the 3d volume can  never vanish on solutions of the equation of motion (as long as $\cC=\pi_{\phi}/\ka$ does not vanish, i.e. as long as there is matter). Now, having two notion of volumes, we can introduce the associated Hubble factors
\be
H = \frac{\dot{v}}{3v}\; \qquad \text{and} \qquad \tilde{H} = \frac{\dot{\cV}}{3\cV}
\ee
The relation between these two Hubble factors can be obtained from (\ref{vVV}) and reads
\be
H = \tilde{H} \left( \frac{1- \tan^2{\left( \lambda b\right)}}{1+ \tan^2{\left( \lambda b\right)}}\right) = \tilde{H} \left( \frac{1- \lambda^2 B^2}{1+\lambda^2 B^2} \right) =  \tilde{H}\left( \frac{4\pi G- \lambda^2\tilde{H}^2 }{4\pi G+ \lambda^2 \tilde{H}^2} \right)
\ee
where $\tilde{H} = 4\pi G B^2$ follows the standard FLRW classical relation.

Summarizing in a few words, this non-linear redefinition of the FLRW cosmology avoids completely the zero-volume singularity. As we show in more details below in the next section, this further replaces  the initial singularity with a bounce.

\subsection{Effective bouncing dynamics}
\label{B3}

We can now focus on the effective dynamics and compute the classical cosmological trajectories induced by the regularized Hamiltonian $\cH$. It will deviate from the usual FLRW trajectories and replace the singularity by an effective bouncing cosmology.

Once more, we can introduce two different notions of energy density based on the two notion of volume, such that the matter density and its regularized version\footnote{The energy density $\tilde{\rho}$ corresponds to the energy density of classical FLRW cosmology. If we introduced the pressure $\tilde{P} = - \partial_{\cV}\left( \tilde{\rho} \cV\right)$, then each satisfy the classical Friedman equations $\tilde{H}^2 = \frac{8\pi G}{3} \tilde{\rho}$ and $\dot{\tilde{H}} = - 4\pi G \left( \tilde{\rho} + \tilde{P}\right)$.} read
\be
\label{rho}
\rho \equiv \frac{\pi^2_{\phi}}{2 a^6V_{\circ}^2} =  \f12\frac{\pi^2_{\phi}}{v^2}\;,   \qquad \tilde{\rho} \equiv \frac{\pi^2_{\phi}}{2 a^6V_{\circ}^2 \cos^4{\left( \lambda b\right)}} =  \f12\frac{\pi^2_{\phi}}{\cV^2} 
\ee
while the associated pressure is defined as 
\be
P = - \partial_{v}\left( \rho v\right)\;. 
\ee Hence, by definition, the energy densities $\rho$ and $\tilde{\rho}$ scale respectively as $v^{-2}$ and $\cV^{-2}$. A direct consequence is that $\rho$ satisfies the standard continuity equation:
\be
\dot{\rho} + 3H \left( \rho + P\right) =0\,.
\ee
\footnotetext{
We write, as long as $\lambda b \in [0,\f\pi4]$:
\begin{align}
\cos^2{\lambda b}
=
\frac{1}{2} \left( 1+ \cos{(2\lambda b)}\right)
=
\frac{1}{2}
\left( 1+ \sqrt{ 1 -\sin^2{(2\lambda b)}}\right)
= \frac{1}{2}
\left( 1+ \sqrt{1 - \frac{\rho}{\rho_c}}\right)
\,.
\end{align}
If $\lambda b$ is larger than $\f\pi 4$, then we need to switch the sign in front of the square-root. Actually, for standard trajectories $\lambda b$ goes to 0 in the classical regime at infinity for large volume and reaches its maximal value $\lambda b=\f\pi4$ at the bounce. It would nevertheless be interesting to analyze non-standard trajectories to investigate their physical meaning.}

Let us now express the $\sl(2,\R)$-covariant Hamiltonian scalar constraint \eqref{Hilqc} in terms of the matter density $\rho$:
\begin{align}
\label{heffdyn}
\cH[N]
=
\f{N}2\left(
\f{\pi_{\phi}^2}{ v \cos^2\lambda b}-\ka^2v\f{\sin^2\lambda b}{\lambda^2}
\right)
=
\f{Nv}{ \cos^2\lambda b}\Big{[}\rho-\rho_{c}\sin^2(2\lambda b)\Big{]}
\,,
\end{align}
where we recall that $\ka^2=12\pi G$ and where we have introduced the {\it critical density}:
\be
\rho_{c}^{-1}
\equiv
\f{8\lambda^2}{\ka^2}
=
\f{(2\lambda)^2}{6\pi G}
\,.
\ee
Solving this constraint (\ref{heffdyn}) allows to express the matter density in terms of  $b$:
\be
\label{rhoclassical}
\cH=0
\Rightarrow
\rho
=
\rho_{c}\sin^2(2\lambda b)
\,.
\ee
We will explain below, when analyzing the equations of motion,  in which sense the value  of the density $\rho_{c}$ is critical.


Setting the lapse $N=1$, it is straightforward to derive the modified Friedman equations for the Hubble parameter $H$ describing the cosmic evolution with respect to the cosmic time $t$:
 \begin{align}
 \label{Friedmod1}
H^2 
& =
\frac{8\pi G}{3} \rho \left[ 1 - \frac{\trho}{4\rho_c}\right]^2
\\
 \label{Friedmod2}
\dot{H} & = - 4\pi G \left( \rho + P\right) \left( 1- \frac{\tilde{\rho}}{4\rho_c}\right) \left[ 1 - \frac{\tilde{\rho}}{4\rho_c} - \frac{\tilde{\rho}}{2\rho_c \sqrt{1- \frac{\rho}{\rho_c}}} \right]
\end{align}
Combining these two equations gives back the classical continuity equation $\dot{\rho} + 3H \left( \rho + P\right) =0$, which indicates that these modified Friedman equations are valid for an arbitrary perfect fluid beyond the present case of a massless scalar field.

The LQC corrections to the Friedman equations are encoded in the factors in brackets $[\dots]$ above in \eqref{Friedmod1} and \eqref{Friedmod2} and become trivial in the small density limit corresponding to the large volume regime. If we remove these corrections, we recover as expected the standard Friedman equations for a classical FLRW cosmology:
\be
\left|
\begin{array}{lcl}
H^2 
&=& \frac{8\pi G}{3} \rho
\\
\dot{H}
&=& -4\pi G \left( \rho + P\right)
\end{array}
\right.
\quad\Rightarrow\quad
\dot{H}+H^2= - \frac{4\pi G}{3} \left( \rho + 3 P\right)
\,.
\ee
The LQC correction lead to an important deviation from the standard dynamics when the density grows. Indeed, when the renormalized density $\trho$ reaches $4\rho_{c}$, which corresponds in terms of the original matter density, this simply corresponds to the critical density $\rho=\rho_{c}$. At this density threshold, the Hubble factor and its first derivative are given by
\be
H\big{|}_{\rho=\rho_{c}}=0\;, \qquad \dot{H}\big{|}_{\rho\rightarrow\rho_{c}} \sim 16\pi G \left( \rho + P\right) \frac{\rho}{\rho_c} 
\ee
This means that  the time-derivative $\dot{H}|_{\rho_{c}} $ is positive which ensures a bounce. At this critical point, we have:
\be
\cos^2 \lambda b=\sin^2\lambda b =\f12
\,,\qquad
\lambda b =\f\pi 4\,.
\ee
We can also compute the critical volume:
\be
v=\f{\pi_{\phi}}{\sqrt{2\rho}}
\quad\Rightarrow\quad
v_{c}=2\lambda\,\f{\pi_{\phi}}{\ka}
\,.
\ee
It is also interesting to check the values of the modified volume $\cV$ and its conjugate variable $B$ at the critical density:
\be
\left|
\begin{array}{lcl}
\cV
&=&
v\cos^2\lambda b
\\
B
&=&
\lambda^{-1}\tan\lambda b
\end{array}
\right.
\quad\Rightarrow\quad
\left|
\begin{array}{lcl}
\cV_{c}
&=&
\f12v_{c}
=\lambda\,\f{\pi_{\phi}}{\ka}
\\
B
&=&
\lambda^{-1}
\end{array}
\right.
\,,
\ee
with thus $\cC_{c}=\cV_{c}B_{c}=\ka^{-1}\pi_{\phi}$ as expected on cosmological trajectories (since $\cC$ is a constant of motion always given by $\pm\ka^{-1}\pi_{\phi}$ as soon as the Hamiltonian constraint is satisfied $\cH=0$). The critical value of the first derivative of the Hubble factor takes then the value
\be
\dot{H}\big{|}_{\rho\rightarrow\rho_{c}} = 32 \rho_c = \frac{\kappa^2}{3\lambda^2}
\ee

\bigskip
Another simple way to derive this bouncing dynamics is the crucial remark that this version of LQC is a non-linear realization of the usual FLRW cosmology, as showed earlier in section \ref{nonlinear}. The volume $v$ of the $3d$ hypersurface can be computed from the regularized volume $\cV$  through the relation
\be
\label{vV}
v=\cV+\f{\lambda^2\cC^2}{\cV}\,,
\ee
where $\cV$ follows the usual classical Friedman equation. The factor $\cC^2$ is a constant of motion equal to $\ka^{-2}\pi_{\phi}^2$.
Although this regularized volume $\cV$ is allowed to run from 0 to $+\infty$, the 3d volume $v$ can never vanish. In fact, we easily compute its minimal value allowed by this relation:
\be
\label{Vmin}
v_{min}=2\lambda\f{\pi_{\phi}}{\ka}
\quad\textrm{obtained for}\quad
\cV_{min}=\lambda|\cC|=\lambda\f{\pi_{\phi}}{\ka}
\,.
\ee
This gives exactly the values of the volume at the critical densities. Notice that the relation (\ref{vV}) completely encodes the bouncing dynamics of our effective quantum cosmology. In the next subsection, we will show that, while this crucial relation actually receives higher order quantum corrections from the back-reaction of the higher moment of the wave function in the deep Planckian regime, such higher order corrections remain always negligible for sufficiently semi-classical quantum states at large volume. For this reason, this relation remains a good effective description of the bouncing dynamics, even in the deep quantum regime.

\medskip

Moreover, the bouncing dynamics equation \eqref{vV} is invariant under a surprising transformation, exchanging small and large (regularized) volumes $\cV\leftrightarrow \lambda^2\cC^2/\cV$, while $\lambda^2\cC^2=\lambda^2\pi_{\phi}^2/\ka^2$ is a constant of motion. This has two important consequences:
\begin{itemize}

\item The evolution of the regularized volume $\cV$ from $\cV_{min}=\lambda{\pi_{\phi}}/{\ka}$ to $+\infty$ maps onto the after-bounce expanding phase  for the 3D volume $v$, while the trans-planckian regime $\cV\in]0, \cV_{min}]$ maps onto the whole before-bounce contracting phase for the 3D volume $v$. Furthermore, the invariance of the dynamics under inversion of $\cV$ means that the near-singularity dynamics for $\cV\rightarrow 0$ is dual to the large volume regime of the theory $\cV\rightarrow+\infty$. This {\it UV/IR duality} implies that the near-singularity dynamics is as semi-classical as the large volume dynamics: indeed both contracting and expanding phases are semi-classical for large 3D volume $v$ and the only true deep quantum regime is at the bounce for $v\sim v_{min}$ close to its minimal value and $\cV\sim \cV_{min}$ close to its turning point. Such UV/IR mixing is reminiscent of the phenomenology of non-commutative field theories (e.g. \cite{Buric:2009ss}) and of the T-duality of string theory.

\item While the actual 3D volume $v$ is not monotonous during the cosmological evolution and bounces, so that it cannot be used as a legitimate time parameter describing the whole evolution of the universe, the regularized volume $\cV$ completely sidesteps this issue and evolves monotonously from 0 to $+\infty$, encoding the whole dyamics from the contracting phase through the bounce to the expanding phase.

\end{itemize}

\medskip

Additionally, the non-linear mapping between the modified LQC cosmology and the standard FLRW cosmology allows for a finer description of the bouncing trajectories. Using the deparametrized picture, describing the evolution of the trajectories with respect to the scalar field $\phi$, the regularized volume follows the usual trajectory:
\begin{align}
\label{Vtraj}
\cV[\phi]=\cV_{0}\,e^{\eps \ka (\phi-\phi_{0})}\,,
\end{align}
with the sign $\eps=\pm$ signaling either an expanding or contracting  trajectory.
The  3d volume $v(\phi)$ obtained by \eqref{vV} necessarily mixes  both sectors:
\begin{align}
\label{vtraj}
v[\phi]= \cV_{0}\, e^{+\eps\ka (\phi-\phi_{0})} + \f{\lambda^2\cC^2}{\cV_{0}} e^{- \eps\ka(\phi-\phi_{0})}\,.
\end{align}
Whatever sign $\eps$ one chooses, we get the same behavior: as $\phi$ increases, the universe evolves from a contracting phase to an expanding phase through a big bounce at its minimal value of the volume $v_{min}=v_{c}$. In fact, the choice of branch $\eps=\pm$ is irrelevant and both signs describe the same cosmological trajectories.

More precisely, the standard FLRW trajectories as written above in \eqref{Vtraj} are a priori labeled by the sign $\eps$, the choice of time origin $\phi_{0}$ and the initial volume $\cV_{0}$ at $\phi=\phi_{0}$. Actually once $\eps$ is chosen, only the value of $\cV_{0}e^{-\eps\ka\phi_{0}}$ matters and it is indeed a constant of motion:
\be
\cV_{0}e^{-\eps\ka\phi_{0}}=\cV(\phi)\,e^{-\eps \ka \phi}
\,.
\ee
And we can label a trajectory by the sign $\eps$ and the value of this observable $\cO\in\R_{+}$:
\be
\cV^{(\eps,\cO)}[\phi]
=
\cO\,e^{\eps \ka \phi}\,.
\ee
Finally, even the value of $\cO$ does not matter since the curves $\cV^{(\eps,\cO)}[\phi]$ for two different values of $\cO$ are still the same. The functions only differ by the choice of time origin $\phi_{0}$:
\be
\cV^{(\eps,\cO)}[\phi]
=
\cV^{(\eps,\cO')}\left[
\phi+\f\eps{\ka}\ln\f{\cO}{\cO'}
\right]
\,.
\ee
In the present context, the specific value of $\phi$ is not a physical observable. More precisely, since we are studying a massless scalar field, we do not have a potential depending on $\phi$ and  its conjugate momentum $\pi_{\phi}$ is a physical observable. We do not have a physical way to determine the value of the scalar field and the choice of the origin time $\phi_{0}$ is arbitrary. In a more general context, once the scalar field acquires a mass or a potential, the analysis of the constants of motion and relevant Dirac observables will have to be revisited. But for now, the deparametrized physical trajectories for the volume $\cV^{(\eps,\cO)}[\phi]$ only depend on $\eps$ in the end. Then we deduce the trajectory for the conjugate variable $B$ from the value of the constant of motion $\cC=\cV B$:
\be
B^{(\eps,\cO,\cC)}[\phi]
=
\f{\cC}{\cV^{(\eps,\cO)}[\phi]}
=
\f\cC\cO\,e^{-\eps \ka \phi}
\,.
\ee

\begin{figure}[h]
\centering
\includegraphics[scale= .6]{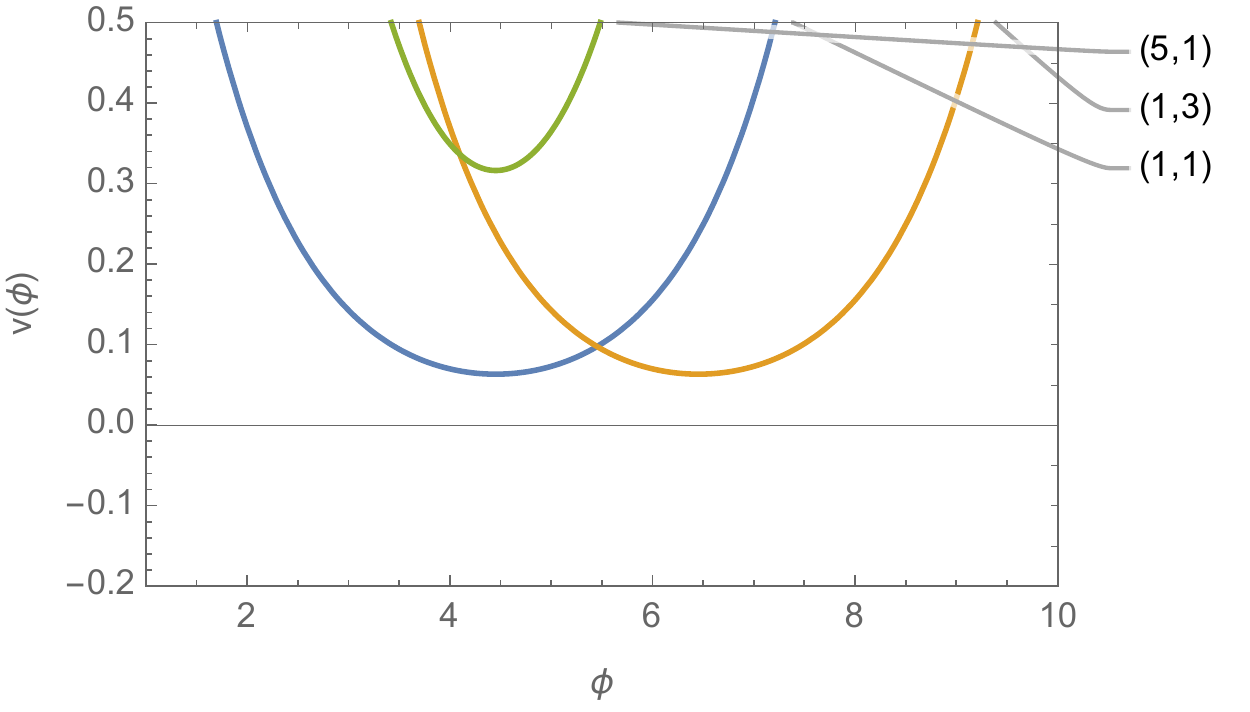}\hskip1cm
\caption{Evolution of the deparametrized volume $v(\phi)$ for different values of the couple of free constants $\left(\cV_0, \phi_0\right)$ and for $\cC=1$ and $\lambda= 0.1$. We observe that the values of $\left(\cV_0, \phi_0\right)$ do not change qualitatively the bounce and the dynamics is symmetric w.r.t the expanding and contracting branch. The blue, green and orange curves  correspond respectively to $\left(\cV_0=1, \phi_0=1\right)$, $\left(\cV_0=5, \phi_0=1\right)$, $\left(\cV_0=1, \phi_0=3\right)$. A shift in $\cV_0$ leads to a shift of the bouncing point towards larger volume (green curved), while a shift in $\phi_0$ leads to a bounce at a different scalar field time (orange curved).
\label{fig:bounce} }
\end{figure}

Now turning to the new bouncing trajectories \eqref{vtraj}, we follow the same line of reasoning. What interests us especially is that the contributions of the two branches for $e^{\pm\eps\ka \phi}$ do not seem to carry the same weight and that the evolution of the volume does not seem to be symmetric as $\phi$ goes to $+\infty$ or to $-\infty$. Moreover, this would seem that the choice of sign $\eps=\pm$ would be relevant, which would mean that we could distinguish a contracting and expanding trajectory despite the singularity being resolved into a bounce. This is actually not true: one can completely re-absorb a switch of sign $\eps \rightarrow -\eps$ by a simple change of time origin $\phi_{0}$, i.e. by a simple translation in $\phi$ without changing the values of the other constants of motion determining the trajectory:
\be
v^{\eps,\cO,\cC}[\phi]
=
\cO\, e^{+\eps\ka \phi} + \f{\lambda^2\cC^2}{\cO} e^{- \eps\ka\phi}
=
v^{-\eps,\cO,\cC}\left[\phi+\f{2\eps}{\ka}\ln\f\cO{\lambda|\cC|}\right]
\ee
Finally, as we can see on figure \ref{fig:bounce}, the trajectory $v^{\eps,\cO,\cC}[\phi]$ is actually completely symmetric around the minimal value of the volume, i.e. under $\phi \rightarrow 2\phi_{\min}-\phi$ with $\phi_{min}$ the value of the scalar field at the critical density $v[\phi_{min}]=v_{min}=v_{c}$.

\medskip

To conclude the analysis of the bouncing cosmological dynamics at the classical level, we would like to emphasize that it should be thought as an effective cosmological dynamics taking into account quantum gravity corrections to the cosmological evolution. The logic is that we start with standard cosmology, quantize gravity, look at the quantum dynamics of cosmological states and derive the  effective dynamics for semi-classical cosmological wave packets.
Indeed, one can view \eqref{heffdyn} as the expectation value of the polymer scalar constraint on semi-classical states. A priori, these states can have a large spread and the expectation value of the scalar constraint could be drastically modified by higher order corrections. Whether or not the above corrections capture all the relevant quantum corrections requires a more detailed analysis which we discuss belows in section-\ref{33}. This will provide a consistency check that the effective dynamics considered here is not modified by higher order corrections and is indeed a good approximation to the bouncing dynamics even in the deep Planckian regime.

\subsection{Robustness of the effective equations}
\label{33}

In section-\ref{B3}, we have computed the effective dynamics of our bouncing cosmology starting from the effective polymer scalar constraint (\ref{heffdyn}). A central question we would like to study now is whether the effective Friedman equations obtained from (\ref{heffdyn}) capture all the relevant quantum corrections ? Indeed, (\ref{heffdyn}) should be understood as the expectation value of the operator version of the scalar constraint on some quantum states which are sufficiently semi-classical at large volume. A priori, such states can experience a large spreading in the deep quantum regime and the higher order moments can back-react on the expression (\ref{heffdyn}), modifying drastically the effective corrections introduced by the naive regularization. In that case, one could not trust anymore the effective dynamics derived in section-\ref{B3}. A careful analysis of the dynamics of the higher order moments of semi-classical states is therefore necessary to justify a posteriori the effective equations. In the context of standard LQC, this analysis was performed in \cite{Bojowald:2007bg, Bojowald:2014uaa, Bojowald:2006gr, Bojowald:2009jk}, showing the robustness of the effective dynamics against the back-reaction of higher order moments of the wave function. See also \cite{Kaminski:2010yz, MenaMarugan:2011me} for a different approach.

Let us briefly summarize the formalism used in \cite{Bojowald:2007bg, Bojowald:2014uaa, Bojowald:2006gr, Bojowald:2009jk}. For additional details, see \cite{Bojowald:2001ae, Bojowald:2009zzc, Bojowald:2009jj, Brizuela:2014cfa, Baytas:2018gbu}.
Consider a quantum system with basic operators $[\hat{q}, \hat{p}]= i \hbar$ and an hamiltonian $\hat{H}(\hat{q}, \hat{p})$. This system can be described by a infinite dimensional quantum phase space in which the canonical variables are given by the expectation value of the basic operators as well as the expectation values of all the higher order fluctuations and correlations associated to these background operators. The expectation values of fluctuations and correlations of any order can be compactly written as
\be
G^{q^a p^b} := \la \left(  \hat{q}  - \la \hat{q}\ra\right)^a \left(\hat{p} - \la \hat{p} \ra \right)^b \ra_{\text{Weyl}}
\ee
which is non zero only for $\ a+ b \geqslant 2$ and the subscript means the Weyl ordering. These expectations values can be understood as quantum variables since they do not have any counterpart in the classical case. The quantum trajectories can therefore explore additional dimensions in the quantum phase space than their classical counterpart which propagates in a finite two dimensional classical phase space. This provides an efficient way to capture the quantum deviations of a given system compare to its classical dynamics. Now, in order to endow the phase space with a symplectic structure, one can define the Poisson bracket between any two quantum phase space functions $\la \hat{X}\ra$ and $\la \hat{Y}\ra$ as
\begin{align}
\{ \la \hat{X}\ra, \la \hat{Y}\ra\} := \frac{1}{i\hbar} \la [ \hat{X},  \hat{Y}]\ra
\end{align}
This allows to compute the Poisson bracket between any expectation values in this quantum phase space. Thanks to this formalism, one works directly at the level of the expectation values without solving explicitly for the wave function.
Instead, a state is determined by the value of the basic operators and all its higher order moments. Notice that this formalism is independent of the representation used to solve the quantum theory, and escapes several crucial representation-dependent difficulties such as the choice of inner product as well as the normalization of the wave function. Such quantum phase space formalism has proven to be a powerful tool to extract semi-classical results from quantum theories, both in quantum cosmology \cite{Bojowald:2010qm, Brizuela:2011aa, Bojowald:2014qha, Brizuela:2019elx} and in more complicated quantum systems \cite{Bojowald:2012vp, Bojowald:2014fca, Baytas:2018wiq}. Finally, let us point that for a constrained system, the basic operators and the infinite set of higher moments, which are the canonical variables of the quantum phase space, are constrained by an infinitely set of constraints, as well as by a set of uncertainty relations reducing the number of independent canonical variables as well as the region that can be explored in the quantum phase space. See \cite{Bojowald:2007bg, Bojowald:2014uaa} for details.
As a last point, we stress that any quantum phase space function which depends on the expectations values of the basic operators as well as on the expectation values of all the higher moments can be Taylor expanded in term of the moment as follow
\begin{align}
\la \hat{X} \ra = X (\la \hat{q}\ra , \la \hat{p}\ra ) + \sum^{+\infty}_{a, b =1} \frac{1}{a! b!} \frac{\partial^{a+b} X}{\partial^a \la \hat{q}\ra \partial^b \la \hat{p} \ra} G^{q^a p^b}
\end{align}
with $a+b \geqslant 2$. If all the terms are included, this expression is exact, but one can also truncate this expansion and work only at a given order in term of the moments, which turn out to be useful approximation to extract semi-classical results.

 In the following, we shall use the same formalism to discuss the robustness of our new effective equation. We will work at the level of the deparametrized dynamics, i.e in term of the scalar field time. Our bouncing dynamics is completly encoded in the relation (\ref{vV}) between the physical volume of the universe $v$ and the regularized volume $V$, which is true at all time at the effective level. Working at the level of the expectation value, our goal is to show that, for quantum states which are sufficiently semi-classical at large volume, the relation (\ref{vV}) receives higher order corrections from the higher moments of the wave fonction which are always negligeable. Following the usual procedure, we only consider second order moments of the wave function.

As a first step, let us compute the dynamics of our $\sl(2,\mathbb{R})$ generators. The hamiltonian of the deparametrized dynamics is given by $\hat{\pi}_{\phi} = \epsilon  \kappa \hat{C} = \epsilon \kappa \hat{\cK}_y$ with $\epsilon = \pm 1$. One can choose arbitrarily the expanding or contracting branch, the only difference being in the definition of the semi-classical regime, i.e the regime of large volume. Using the formalism detailed in  \cite{Bojowald:2007bg, Bojowald:2014uaa, Bojowald:2006gr, Bojowald:2009jk}, one obtains
\begin{align}
\frac{d \la \hat{\cJ}_z \ra}{d\eta} = \frac{1}{i\hbar} \la [\hat{\cJ}_z, \hat{\cK}_y]\ra = - \la \hat{\cK}_x\ra \qquad \text{and} \qquad \frac{d \la \hat{\cK}_x \ra}{d\eta} = - \la \hat{\cJ}_z \ra \qquad \frac{d \la \hat{\cK}_y \ra}{d \eta} =  0
\end{align}
where $\eta = \epsilon \sqrt{12\pi G} \phi$ is the scalar field time. This allows to obtain the dynamics of the expectation values of our CVH generators straitforwardly which read
\begin{align}
\frac{d \la \hat{V} \ra}{d\eta} = -  \kappa \la \hat{V}\ra \qquad \frac{d \la \hat{\cC} \ra}{d\eta} = 0  \qquad \frac{d \la \hat{\cH} \ra}{d \eta} \simeq  0
\end{align}
which implies that 
\be
\la \hat{V} \ra (\eta)= c_1 e^{- \eta} \qquad \la \hat{C} \ra(\eta) = c_2
\ee
with $c_1$ and $c_2$ two constants to be fixed latter. If one picks up the negative branch, i.e $\epsilon = -1$, the large volume limit, and thus the semi-classical regime corresponds to $\eta = - \sqrt{12\pi G} \phi \rightarrow - \infty$ and thus $\phi \rightarrow + \infty$. Notice also that for $\tau \rightarrow 0$, the expectation value of the regularized volume can be arbitrarily closed to zero\footnote{This is reminiscent of the failure of the Wheeler-De Witt quantization, based on the Schrodinger representation, to resolve the past singularity.}, which is expected since it follows the classical Friedman equation, as one can see from (\ref{nonlinearFRW}) or (\ref{Vtraj}). 

Let us introduce now the second order fluctuations of the operators $\hat{V}$ and $\hat{C}$. They corresponds to two fluctuations and one correlation given by
\begin{align}
& G^{VV} = \la \hat{V}^2\ra - \la \hat{V}\ra^2 \qquad G^{CC} = \la \hat{C}^2\ra - \la \hat{C}\ra^2 \\
& G^{VC}= \frac{1}{2}\la \hat{V} \hat{C} + \hat{C} \hat{V}  \ra - \la \hat{V}\ra \la \hat{C}\ra
\end{align}
Their dynamics can be computed as well and read
\be
\frac{G^{VV}}{d\eta} = - 2 G^{VV}  \qquad \frac{G^{CC} }{d\eta} = 0 \qquad \frac{G^{VC} }{d\eta} = - G^{VC}
\ee
which is solved for
\be
G^{VV} (\eta) = c_3 e^{-2 \eta} \qquad G^{CC}(\eta) = c_4 \qquad G^{VC}(\eta) = c_5 e^{- \eta}
\ee
Notice that although the fluctuations are not constant in the scalar field time, the ratios
\be
\label{fluctu}
\frac{ G^{VC}}{\la \hat{V} \ra  \la  \hat{C}\ra} = \frac{c_5}{c_1 c_2},  \qquad  \frac{ G^{VV}}{ \la \hat{V}\ra^2} = \frac{c_3}{c^2_1} \qquad \text{and} \qquad \frac{ G^{CC}}{ \la \hat{C}\ra^2} = \frac{c_4}{c^2_2}
\ee
remain constant during the whole quantum evolution. In particular, if one starts with a semi-classical state at large volume, i.e at $\tau \rightarrow - \infty $, such that its spreading is tiny, one can conclude that 
\be
\frac{c_5}{c_1 c_2} \ll 1, \qquad \frac{c_3}{c^2_1} \ll 1, \qquad \frac{c_4}{c^2_2} \ll 1
\ee
at any time of the cosmic evolution for such state.

We can now investigate how the expectation value of the physical volume of the universe is affected by the higher moments. Indeed, the effective bounce can be described by the relation (\ref{vV}) which is non linear due to the inverse volume term. Therefore, we do expect the higher moments to couple to the background expectation value $\la \hat{V}\ra$ and a priori, this effective relation can be modify by this backreaction. Using a Taylor expansion up to second order in the moment, one can write the expectation value of the physical volume $\la \hat{v} \ra$ as a function of the expectation values $(\la \hat{V} \ra , \la  \hat{C}\ra)$ and of the second order correlations $(G^{VV}, G^{CC}, G^{VC})$ as:
\begin{align}
\la \hat{v} \ra
& \simeq \la \hat{V}\ra  + \frac{\lambda^2 \la \hat{C}\ra^2}{\la \hat{V}\ra } \left( 1 - \frac{2 G^{VC}}{\la \hat{V} \ra  \la  \hat{C}\ra} + \frac{ G^{CC}}{\la \hat{C}\ra^2}  + \frac{ G^{VV}}{2 \la \hat{V}\ra^2} + \cO \left(  \frac{ G^{VVV}}{\la \hat{V}\ra^3}\right) \right)  \nn \\
& = \la \hat{V}\ra  + \frac{\lambda^2 \la \hat{C}\ra^2}{\la \hat{V}\ra } \left( 1 - \frac{2c_5}{c_1 c_2} +  \frac{c_4}{c^2_2} +  \frac{c_3}{2c^2_1} + \cO \left(  \frac{ G^{VVV}}{\la \hat{V}\ra^3}\right) \right) 
\end{align}
The last term which contains the inverse volume term generates as expected a coupling with the higher moments. However, as shown in (\ref{fluctu}), the three last terms involving the second order fluctuations and correlation are constant during the whole evolution. This implies that for sufficiently semi-classical states peaked on a classical volume in the large volume regime, i.e $\eta \rightarrow - \infty$, the higher moments contributions to the effective equation (\ref{vV}) will remain negligible during the whole cosmic evolution. Moreover, the first term linear in the volume is dominant in this semi-classical limit. The relation (\ref{vV}) remains therefore a good approximation, even in the deep planckian regime, when including the second order moments of the wave function. We conclude that one can safely work at the effective level, which provides a good approximation of the quantum bouncing dynamics. 

Finally, let us point that the underlying $\SL(2,\mathbb{R})$ structure of this cosmological model allows to build an elegant coarse-graining of the quantum theory based on group coherent state technics, including the contributions of all the higher moments, as discussed in \cite{Bodendorfer:2018csn}.

\section{Polymer quantization preserving the $\SL(2,\R)$ structure}
\label{polymerquantization}
\label{B2}

Let us turn to the quantization of the theory. There are two natural quantization schemes at our disposal:
\begin{itemize}
\item On the one hand, we can exploit the $\SL(2,\R)$ structure generated by the CVH algebra and straightforwardly quantize the system as a $\SL(2,\R)$ representation. 
\item On the other hand, we can use the standard toolkit of Loop Quantum Cosmology (LQC) based on the polymer quantization. 
\end{itemize}

In this section, we focus on the realization of the polymer quantum theory. We will discuss  the alternative $\SL(2,\R)$  group quantization in the following section \ref{C} and compare it to the polymer scheme. 

\medskip

When performing the polymer quantization of our $\SL(2,\mathbb{R})$-invariant cosmological system, a natural question is how to preserve the conformal symmetry when introducing a regularization involving a fundamental length scale $\lambda$. Indeed, as a fundamental step prior to quantization, one has to regularize the operator $b$, i.e the extrinsic curvature, in order to proceed to the polymer quantization.
This introduces an upper bound on the extrinsic curvature. As in standard loop quantum cosmoogy (LQC), this regularization, crucially combined with the polymer quantization, leads to alternative quantum cosmology  and resolves the big bang cosmology into a big bounce \cite{LQCReview}, mixing the expanding and contracting branch of the classical dynamics through a quantum superposition of the associated quantum geometries.

\subsection{Kinematical quantum states: Revisiting LQC}

Let us now move to the quantum level and proceed to the polymer quantization of the new regularized phase space.  and discuss the realization of the conformal $\sl(2,\mathbb{R})$ symmetry at the quantum level in this quantization scheme. Let us briefly recall the basis of the polymer quantization before applying it to our cosmological system. 

Consider the exponentiated operators $\widehat{U}_{\lambda}$ and $\widehat{W}_{\mu}$ defined as
\begin{align}
\widehat{U}_{\lambda} : =
e^{i \lambda \hat{b}}
\qquad
\widehat{W}_{\mu} : =
e^{ i \mu \hat{v}}
\end{align} 
where  $(\lambda, \mu)$ are two scales with dimensions $[\lambda] = [\mu]^{-1} = L^{3}$.
The standard commutation relations for the operators $\hat{b}$ and $\hat{v}$ become 
\begin{align}
\label{weyl}
\widehat{U}_{\lambda} \widehat{W}_{\mu} = e^{i\lambda \mu} \widehat{W}_{\mu} \widehat{U}_{\lambda}\qquad \widehat{U}_{\lambda_1} \widehat{U}_{\lambda_2} = \widehat{U}_{\lambda_1 + \lambda_2} \qquad \widehat{W}_{\mu_1} \widehat{W}_{\mu_2} = \widehat{W}_{\mu_1+\mu_2}
\end{align}
which form the Weyl algebra. The polymer quantization is a realization of this algebra of operators on a non-separable Hilbert space $\cH_{\text{Pol}}$. Working in the volume representation, the polymer Hilbert space is spanned by mutually orthogonal vectors $| v \ra$, with $v \in \mathbb{R}$ such that
\be
\label{scal}
\la v | v' \ra_{\text{Pol}} = \delta_{v,v'}
\ee
A wave function in $\cH_{\text{Pol}}$ can be decomposed in this basis as
\begin{align}
\Psi (v) = \sum_{v \in \mathbb{R}} c (v) | v \ra \qquad \text{such that} \qquad  \sum_{v \in \mathbb{R}} |c (v)|^2 < \infty
\end{align}
The operators $\widehat{U}_{\lambda}$ and $\widehat{V}_{\mu}$ act on a basis ket $| v \ra$ as
\begin{align}
\widehat{U}_{\lambda} | v \ra = | v + \lambda \ra \qquad \widehat{W}_{\mu} = e^{i\mu v} | v \ra
\end{align}
and satisfy $\widehat{U}^{\dagger}_{\lambda} = \widehat{U}_{-\lambda}$ and $\widehat{W}^{\dagger}_{\mu} = \widehat{W}_{-\mu}$.
The crucial property descending from the choice of Hilbert space and scalar product (\ref{scal}) is that the mapping of the real line on itself through
\begin{align}
\zeta: & \; \mathbb{R} \rightarrow \mathbb{R} \\
& \; \lambda \rightarrow \la v | \widehat{U}_{\lambda}| v \ra_{\text{Pol}} = \delta_{v, v+\lambda}
\end{align} 
is not continuous. Indeed, since  $\lambda$ can be taken very small but is always different from 0, the scalar products $\la v |  v+ \lambda \ra_{\text{Pol}} =0$ always vanish. This implies in turn that the infinitesimal version of the operator $\widehat{U}_{\lambda}$ is not well defined anymore, and the operator $\hat{b}$ does not exist in this polymer representation. This discontinuous property of the operator $\widehat{U}_{\lambda}$ allows to encode a simple notion of discrete geometry in quantum cosmology, an ingredient which is not present in the standard WdW quantization, leading therefore to an inequivalent lattice-like quantum theory of the universe. The standard WdW quantum cosmology is recovered upon coarse-graining the lattice structure \cite{Fredenhagen:2006wp, Corichi:2007tf}. More details on polymer quantum mechanics can be found in \cite{Corichi:2006qf, ChaconAcosta:2011vv, Date:2012gf, Majumder:2012qy, Kunstatter:2012np, Gorji:2015kta, Flores-Gonzalez:2013zuk, Morales-Tecotl:2016ijb, Berra-Montiel:2018pkz}. Finally, the multiplicative operator $\widehat{W}_{\mu}$ remains continuous w.r.t to the scale $\mu$, and one can instead work with its infinitesimal version, namely $\hat{v}$, which remains well defined. The uniqueness property of this quantum mechanics have been shown in \cite{Engle:2016zac}. See also \cite{Ashtekar:2012cm} for a discussion on this aspect.

\bigskip

We apply now the polymer quantization to our cosmological model. The commutation relation of the gravitational and matter degrees of freedom read
\be
\label{commutrel}
[\widehat{U}_{\pm \lambda}, \; \hat{v}] = \mp \; \lambda\;  \widehat{U}_{\pm \lambda} \qquad [\hat{\phi}, \hat{\pi}_{\phi}]= i \mathds{1}
\ee
The matter and gravitational sectors are quantized separately.  The quantum operator act on the wave-functions $\Psi(\phi, v)$ as
\begin{align}
\label{exp}
\widehat{U}_{\pm \lambda} \, \Psi(\phi, v) &=  \Psi(\phi, v \pm \lambda) \qquad  \hv \,\Psi(\phi, v) = v\, \Psi(\phi, v) \\
\hat{\pi}_{\phi} \, \Psi(\phi, v) &= - i \frac{\partial}{\partial\phi}\;  \Psi(\phi, v ) \qquad \hat{\phi} \, \Psi(\phi, v) = \phi \;  \Psi(\phi, v )
\end{align}
The consequence of working in the polymer representation is that we obtain superselection sectors for the gravitational sector. Acting with operators generated by $\widehat{U}_{\pm \lambda} $ and $\hv$ only create integer shifts in the volume. Thus the volume takes discrete real values given by 
\be
v_n = \epsilon \pm  \lambda n
\ee
with $n \in \mathbb{Z}$ and a fixed offset $\epsilon \in [0,\lambda[$. Different values of $\eps$ define superselection sectors. This is the lattice-like structure of LQC. The matter degrees of freedom are quantized in the standard Schr\" odinger representation while the gravitational d.o.f are quantized following the polymer scheme. The resulting Hilbert space is  a tensor product of the matter and gravitational Hilbert spaces ${\mathbf H}_m \otimes {\mathbf H}_g$.
Quantum states are wave-functions  $\Psi(\phi, v)$satisfying a $L^2$ normalization condition:
\begin{align}
\sum_{v \in \mathbb{R}} \int d\phi |\Psi(\phi, v)|^2 < + \infty
\,.
\end{align}
Here we  focus on the choice\footnotemark{} $\eps=0$.
\footnotetext{
Since we would like to restrict physically to positive values of the volume $v$, a natural requirement is to impose the parity of the wave-function $\Psi(\phi,-v)=\Psi(\phi,v)$. This necessarily implies that the spectrum of $v$ be symmetric, which selects the two special values $\epsilon=0$ and $\epsilon=\f\lambda 2$. 
}
Let us quantize the CVH operators in their new regularized version, as given in \eqref{regCVH1} and \eqref{Hilqc}. We split the Hamiltonian constraint into the matter contribution in $\pi_{\phi}^2$ containing the inverse volume factor and the purely gravitational contribution $\cH_{g}$:
\be
\cH_{g}=-\f{\ka^2v\sin^2\lambda b}{2\lambda^2}
=\f{\ka^2\,v}{4\lambda^2}\big{[}\cos(2\lambda b)-1\big{]}
\,.
\nn
\ee
We recall the regularized definition of the dilation generator and of the volume:
\be
\cC=\f{v\sin (2\lambda b)}{2\lambda}
\,,\quad
\cV=v\cos^2\lambda b
=\f{v}2\big{[}1+\cos(2\lambda b)\big{]}
\,.
\ee
In order to realize an $\sl(2,\mathbb{R})$ polymer quantization, we need to find a factor ordering of these operators such that they are self-adjoint and reproduce the $\sl(2,\mathbb{R})$ commutations relations of the classical vacuum phase space (i.e. the purely gravitational sector), namely
\begin{align}
\label{sl2R-puregravity}
[\widehat{\cV}, \widehat{\cH}_g]  = i\ka^2 \widehat{\cC} \;\;\;\;\;\; [\widehat{\cC}, \widehat{\cH}_{g}]  = - i \widehat{\cH}_g \;\;\;\;\;\; [\widehat{\cC}, \widehat{\cV}]  = i \widehat{\cV}
\end{align}
This symmetry criteria allows to select a factor-ordering choice, in which the CVH operators are quantized by splitting the $v$ factor into two square-root operators $\sqrt{|v|}$ inserted on both side of the operators in order to ensure that they are self-adjoint:
\begin{align}
\widehat{\cV}\, \Psi (\phi, v_n)
& = \frac{1}{4}  \sqrt{|v|} \left( \widehat{U}_{+2\lambda}  + \widehat{U}_{-2\lambda}  + 2 \right) \sqrt{|v|}\,  \Psi(\phi, v_n )
\,\\
\widehat{\cC}\, \Psi (\phi, v_n)
& =
\frac{-i}{4\lambda } \sqrt{|v|} \left(\widehat{U}_{+2\lambda}  - \widehat{U}_{-2\lambda}  \right) \sqrt{|v|}\,  \Psi(\phi, v_n )
 \\
\label{h}
\widehat{\cH}_g\, \Psi (\phi, v_n)
& =
\frac{\ka^2}{8\lambda^2 } \sqrt{|v|} \left( \widehat{U}_{+2\lambda}  + \widehat{U}_{-2\lambda}   - 2 \right) \sqrt{|v|} \; \Psi(\phi, v_n )
\end{align}
or equivalently written in the $n$ basis (forgetting the decoupled matter sector), with $v_{n}=n\lambda$:
\begin{align}
\widehat{\cV}\,|n\ra
& = \frac{\lambda}{4}  \Big{[}
2n|n\ra+\sqrt{|n(n+2)|}\,|n+2\ra+\sqrt{|n(n-2)|}\,|n-2\ra
\Big{]}
\,\\
\widehat{\cC}\,|n\ra
& =
\frac{-i}{4} \Big{[}
\sqrt{|n(n+2)|}\,|n+2\ra-\sqrt{|n(n-2)|}\,|n-2\ra
\Big{]}
 \\
\label{h}
\widehat{\cH}_g\,|n\ra
& =
\frac{\ka^2}{8\lambda}
 \Big{[}
-2n|n\ra+\sqrt{|n(n+2)|}\,|n+2\ra+\sqrt{|n(n-2)|}\,|n-2\ra
\Big{]}
\end{align}
Noticing that volume shifts induced by those operators are always in $\pm 2\lambda$, i.e. twice the lattice spacing, we choose for the sake of simplicity to remain in the sector containing the zero volume, i.e. $n\in2\N$. In this case, we can drop the absolute values in the square-root.
The $\sl(2,\mathbb{R})$ symmetry fixes completely the factor-ordering ambiguity in the quantum theory, and provides therefore a stringent criteria to build the quantum theory unambiguously. One can further check that the associated Casimir operator in this polymer realization remains null and does not carry quantum corrections compared to the classical case, namely
\begin{align}
-\widehat{\mathfrak{C}}_{\mathfrak{sl}_{2}}
= 
\widehat{\cC}^2
+
\f1{\ka^2}
\Big{[}
\widehat{\cH}_{g}\widehat{\cV}+\widehat{\cV}\widehat{\cH}_{g}
\Big{]}
=0
\end{align}
Therefore, the gravitational sector of our loop quantized theory preserves the classical $\SL(2,\mathbb{R})$ symmetry and the quantum states defined on the quantum lattice live in a null representation of the $\sl(2,\mathbb{R})$ Lie algebra.

Now, as we would like to solve the Hamiltonian constraint, we also need to take into account the matter contribution to the Hamiltonian. The complication is that it contains an inverse volume factor, which inevitably leads to quantization ambiguities. One way around this issue is to consider the Hamiltonian constraint with a choice of lapse given by the volume, $N=\cV$:
\be
\cH[N] =N\left( \frac{\pi^2_{\phi}}{2 \cV}+ \cH_g \right)
\quad\longrightarrow\quad
\cH[\cV] = \frac{\pi^2_{\phi}}{2}+ \cV\cH_g 
\ee
Quantizing this expression by properly symmetrizing the term $ \cV\cH_g$ gives the following  smeared scalar constraint operator:
\begin{align}
\label{hv}
\widehat{\cH[\cV]}
=
\frac{\hat{\pi}^2_{\phi}}{2 } + \frac{1}{2} \left(  \widehat{\cV}\widehat{\cH}_g+    \widehat{\cH}_g \widehat{\cV}  \right) 
=
\f{\ka^2}2
\Bigg{[}
\frac{\hat{\pi}^2_{\phi}}{\ka^2}  - \widehat{\cC}^2
\Bigg{]}
\,,
\end{align}
where we used the vanishing Casimir condition. This expression is much simpler to handle. Let us underline that this corresponds to a choice of classical lapse; the issue of consistently defining choices and changes of lapse at the quantum level is still an open issue in quantum gravity.

\subsection{Singularity resolution in deparametrized dynamics}

The Hamiltonian constraint $\cH$ given in (\ref{Hilqc}) contains an inverse volume factor (in the matter term), just like the standard LQC Hamiltonian $\cH$. To sidestep this problem, two strategies are commonly adopted: either one  regularizes this operator at (or around) the 0-volume state (by hand or using a trick \`a la Thiemann), or one switches the time variable and works within the deparametrized picture, which corresponds here to fixing the lapse to $N= \cV$. In the following, we shall focus on the deparametrized dynamics. 

Let us thus consider the Hamiltonian constraint operator $\widehat{\cH[\cV]}$ as defined above in \eqref{hv},
\be
\hat{\pi}^2_{\phi} \Psi(\phi, v) 
=
\ka^2\widehat{\cC}^2 \Psi(\phi, v)
\,.
\nn
\ee
We switch to the Fourier basis for the scalar field and decompose the wave-function as
\be
\Psi=\sum_{n\in2\N}\int_{\R}\rd k\, \Psi^k_{n}e^{ik\ka\phi}\,|\phi,n\ra
\,.
\ee
Each mode $k$ corresponds to an eigenvalue $-\ka^2 k^2$ of the operator $\hat{\pi}^2_{\phi} $ and leads to a a second order difference equation:
%
\begin{align}
\label{eqdiff1}
{(4k)^2} \Psi_n^k
= &
 2 n^2 \Psi_n^k-\sqrt{n (n+ 4)} (n + 2) \Psi_{n+4}^k- \sqrt{n (n- 4)} (n - 2 ) \Psi_{n-4}^k
\end{align}
This is our Wheeler-de-Witt equation. This implements a recursion with volume steps $\pm 4\lambda$.

It is possible to switch back to a difference equation with a volume step $\pm 2\lambda$ equal to the volume lattice spacing. We classically consider the two branches of solutions to the Hamiltonian constraint given by $\pi_{\phi}=\eps\ka \cC$ with the sign $\eps=\pm$.
The splitting of the evolution equation in an operator for the expanding branch and one for the contracting branch is made possible because one can easily split between positive and negative frequency in this homogeneous and isotropic case. 
Choosing the positive branch, we quantize the constraint $\pi_{\phi}=+\ka \cC$, which gives the following difference equation:
\be
\label{eqdiff2}
4ik \Psi_n^k
=
\sqrt{n (n+ 2)}  \Psi_{n+2}^k- \sqrt{n (n- 2)}  \Psi_{n-2}^k
\,,
\ee
which indeed implies the previous Wheeler-de-Witt equation \eqref{eqdiff1}. This equation is neat from the $\sl(2,\R)$ point of view, since it simply involves diagonalizing the $\sl_{2}$ generator $K_{y}$ in the null representation corresponding to this polymer quantization. This is readily done using the results and methods of \cite{lindblad}.

A key feature of both the double-step and the  Wheeler-de-Witt equations \eqref{eqdiff1} and \eqref{eqdiff2} is the resolution of the singularity.
Assuming that the matter content is non-trivial $k\ne0$, one obtains for $n=0$ that
\be
\label{0}
\Psi_0 =0
\ee
Thus the zero volume state is forbidden: physical wave function solutions of (\ref{eqdiff1}) (or (\ref{eqdiff2})) have no support on the zero volume eigenstate. The volume operator never vanishes on physical states solving the quantum Hamiltonian constraint. Notice that contrary to other approaches to quantum cosmology, the vanishing of the zero volume state (\ref{0}) is not removed by some ad hoc boundary condition imposed on the wave function at zero volume, but its decoupling from the other states is derived from the quantum dynamics. 
Moreover positive volumes $n> 0$ decouple from  negative volumes $n<0$. The physical Hilbert space is therefore decomposed as
\begin{align}
{\mathbf H} = {\mathbf H}_{-}  \oplus {\mathbf H}_{+}
\,.
\end{align}
We can thus safely restrict ourselves to positive volume states, without fear of quantum transitions from negative to positive volume states.

\section{$\SL(2,\mathbb{R})$ group quantization and Comparison with LQC}
\label{C}

The LQC polymer quantization we presented in the previous section is a mixed scheme from the point of view of the $\SL(2,\R)$ structure. As it was already underlined in \cite{BenAchour:2017qpb}, the $\SL(2,\R)$ symmetry holds for the whole system of gravity plus matter for the FRLW cosmology, equations \eqref{dil} to \eqref{weightH}, and its polymer regularization, equations \eqref{Hilqc} to \eqref{LQC-sl2R},  and it encodes in both scenarii the dynamics  of the theory. However, the standard polymer quantization, as explained in section \ref{polymerquantization} separates gravity from matter and quantizes the theory with a Hilbert space of quantum states defined as tensor products of gravity states times matter states. This only preserves the vacuum $\SL(2,\R)$ structure in the pure gravity sector, as showed in \eqref{sl2R-puregravity}, but does not allow (at least in a transparent way) for the full $\SL(2,\R)$ structure of gravity plus matter to be represent at the quantum level. It is thus natural to introduce a  $\SL(2,\mathbb{R})$ group quantization, which quantizes the full theory of gravity plus matter as a $\SL(2,\R)$ representation, and compare it to the polymer quantization.

\subsection{Gravity plus matter states as a $\SL(2,\R)$ representation}

We focus on the CVH algebra for regularized LQC, formulated as a $\sl(2,\R)$ algebra as written above in \eqref{LQC-CVH} and \eqref{LQC-sl2R}. The $\sl(2,\R)$ generators are given in term of the regularized CVH observables by
\be
\cJ_z = \frac{\cV}{2\sigma\ka^3} - \sigma\ka\cH
\,,\qquad
\cK_x =  \frac{\cV}{2\sigma\ka^3} + \sigma\ka\cH
\,, \qquad
\cK_y = \cC
\ee 
where $\ka^2=12\pi G$ and $\sigma\in\R_{+}$ is an arbitrary real parameter.
$\cV$ and $\cC$ are the regularized volume and dilatation generator given in \eqref{regCVH1}, while $\cH$ is the resulting new regularized Hamiltonian for LQC derived in \eqref{Hilqc}. The $\SL(2,\R)$ Casimir is given by the matter energy:
\begin{align}
\mathfrak{C}
& =
\cJ_{z}^{2}-\cK_{x}^{2}-\cK_{y}^{2}  =
-\f2{\ka^2}\cV\cH-\cC^{2}
=
-\f{\pi_{\phi}^{2}}{\ka^2}
\,.
\label{Casimircl}
\end{align}
As $\pi_{\phi}$ is a Dirac observable, $\{\cH,\pi_{\phi}\}=0$, and thus a constant of motion, the Casimir of the $\sl(2,\R)$ algebra is fixed and we quantize the polymer-regularized cosmology as a $\SL(2,\R)$  representation with negative quadratic Casimir. Such a representation is part of the principal continuous series of $\SL(2,\R)$  representations and is usually referred to as a ``space-like'' representation\footnotemark{}
\footnotetext{
Indeed these irreducible unitary representations from the principal continuous series are obtained using Kirillov's method as the quantization of co-adjoint orbits given by the 2d one-sheet hyperboloids in the 3d Minkowski space-time, that is the orbit of a space-like vector under the 3d Lorentz group. They can be thus be understood as representing a quantum space-like vector.}
These irreducible representations\footnotemark{} are labelled by a real  parameter $s\in\R$ and the corresponding Hilbert space $V^s$ is spanned by the eigenvectors $|s,m\ra$  of the rotation generator  $\hcJ_{z}$ with eigenvalue $m\in\Z$:
\footnotetext{
We point out that we could use an odd representation of $\SL(2,\R)$, with spectrum $m\in\f12+\Z$, which seems to correspond to the LQC superselection sector $\eps=\lambda$. Moreover, if we use representations of the universal cover of $\SL(2,\R)$, the spectrum will get an arbitrary shift $m\in\eta+\Z$ with $\eta\in[0,1[$, thereby reproducing all the LQC  superselection sectors labeled by $\eps$.
}
\begin{align}
\hcJ_{z}\,|s,m\ra & =m\,|s,m\ra \,,\\
\widehat{\mathfrak{C}}
\,|s,m\ra& = j(j+1)\,|s,m\ra
= -\left(s^2+\f14\right)\,|s,m\ra
\qquad\textrm{with}\quad j = - \frac{1}{2} + i s
\,,
\nn
\end{align}
with the scalar product $\la  s ,m|s,m'\ra = \delta_{m,m'}$.
Comparing this with the value of the Casimir at the classical level \eqref{Casimircl}, we identify the representation label to the matter momentum, $s=\ka^{-1}\pi_{\phi}$, and the $+1/4$ term as a quantum correction to the Casimir.
Then the kinematical Hilbert space for gravity plus matter states are $|\Phi\ra = \sum_m \phi_m |s,m\ra$ with the $L^2$ normalization condition $\sum_{m  \in  \Z}  |\phi_{m} |^2 = 1$. 

Notice that for small value of the kinetic energy, i.e $0< \pi_{\phi}^2\kappa^{-2} < 1/4$, the system falls in the complementary serie where the Casimir is given by $ -1/4< \mathfrak{C} < 0$ and where $m\in\Z $. In the following, we shall only work out the quantization corresponding to $\pi^2_{\phi} \kappa^{-2} > 1/4$, which imposes to work with the principal continuous serie.

The boost generators $\hcK_{x}$ and $\hcK_{y}$ create transitions between states with different weights $m$. More precisely,  the raising and lowering operators $\hcK_{\pm}=\hcK_{x}\pm i \hcK_{y}$ act as
\be
\left|
\begin{array}{l}
\hcK_{+} \,| s, m \ra = \beta_{m+1}\, | s, m +1 \ra \\
\hcK_{-}\, |s, m \ra = \beta_{m}\, | s, m - 1 \ra 
\end{array}
\right.
\ee
with
\beq
\forall m\in\Z\,,\quad
\beta_{m}
&=&
\sqrt{m(m-1) - j(j+1)}
=\sqrt{(m+j)(m-j+1)} \nn
 \\
 &=&
\sqrt{m(m-1)+s^2+\f14}\in\R_{+}
\eeq
This gives the $\sl(2,\R)$ Lie algebra commutations and realizes as wanted the quantization of the CVH algebra of observables:
\be
\hcV={\sigma\ka^3}(\hcK_{x}+\hcJ_{z})
\,,\quad
\hcH=\f1{2\sigma\ka}(\hcK_{x}-\hcJ_{z})
\,,\quad
\hcC=\hcK_{y}
\,,
\ee
\be
[\hcJ_{z},\hcK_{x}]=i\hcK_{y}
\,,\quad
[\hcJ_{z},\hcK_{y}]=-i\hcK_{x}
\,,\quad
[\hcK_{x},\hcK_{y}]=-i\hcJ_{z}
\,,
\ee
\be
[\hcC,\hcV]=i\hcV
\,,\qquad
[\hcC,\hcH]=-i\hcH
\,,\qquad
[\hcV,\hcH]=i\ka^2\hcC
\,.
\ee
The key features of this quantization are as follows.
To start with, the dilatation generator (or complexifier) $\cC$ is quantized as the boost generator $\widehat\cK_{y}$, which is a Hermitian operator in the unitary representations of $\SL(2,\R)$, and thus scale transformations are  implemented as unitary $\SL(2,\R)$ transformations acting on the Hilbert space at fixed $s$.
These scale transformations at the quantum level do not affect the universal minimal length scale $\lambda$ introduced as a regulator for the Hamiltonian constraint and act on the Hilbert space of quantum states for a given $\lambda$.
Moreover, the dilatation generator $\cC$ is the deparametrized Hamiltonian generating the evolution of the geometry with respect to the (massless) scalar field, so the deparametrized dynamics is simply derived as the $\SL(2,\R)$ flow generated by a boost.

Then an important difference with the polymeric quantization scheme, as usually used in LQC, lays in the fact that  the operator $\hcJ_{z}$ has a discrete spectrum, diagonalized by the basis states $|s,m\ra$ with eigenvalues $m\in\Z$. Comparing with the polymer quantization presented in the previous section, this operator $\hcJ_{z}$ is closely related to the volume $v$ but is crucially not the volume. Classically, it is expressed by a combination of the volume plus a matter term containing a regularized inverse volume factor:
\be
\cJ_z 
=
\f v{2\sigma \ka^3}\left(
 \cos^2\lambda b +\f{\sigma^2 \ka^6}{\lambda^2}\sin^2\lambda b
\right)
-
\frac{ \sigma\ka\,\pi^2_{\phi}}{2 v \cos^2{(\lambda b)}}
\,,
\ee
where $\sigma$ is the arbitrary free parameter in the mapping between the CVH observables and the $\sl(2,\R)$ basis generators.
If we make the canonical  choice $\sigma \ka^2\lambda=1$, this simplifies to:
\be
\cJ_z  =
\frac{v}{2 \lambda} 
-
\frac{\lambda  \pi^2_{\phi}}{2\ka^2 v \cos^2{(\lambda b)}}
\,.
\ee
%
%
Hence,  $\cJ_{z}$ does contain the volume $v$ as a first term but it further contains  a matter term. This is a systematic feature of the $\SL(2,\R)$ quantization since the $\SL(2,\R)$ symmetry intrinsically mixes the geometry and matter sectors\footnotemark{} of the model.
\footnotetext{
This feature is actually to be expected from a quantum gravity theory, whose is supposed among other things to provide a unified framework for geometry and matter.}
%
%
First,  while the volume operator in polymeric quantization has a discrete spectrum, it is now that the operator  $\hcJ_{z}$ that has a discrete spectrum. Actually we do not have direct access to a volume operator: there is no straightforward proposal for an operator $\hat{v}$ in the $\SL(2,\R)$ quantization scheme, and it is instead this specific combination of volume and regularized inverse volume that becomes the natural observable to work with.
%
%
Second while in polymeric quantization, it was natural to restrict the volume $v$ to positive value with the issue of the singularity being the behavior of the quantum state at $v=0$, there is now a priori nothing special happening\footnotemark{} at $\cJ_z =0$ and no reason, neither physical nor mathematical, to consider only states with $m\in\N$. In fact, while $v \rightarrow + \infty$ corresponds to $\cJ_z \rightarrow + \infty$, the singularity $v \rightarrow 0$ is mapped to $\cJ_z \rightarrow - \infty$.
\footnotetext{Moreover, the equation $\cJ_z =0$ translates into different conditions on $v$ and $b$ depending on the value of the parameter $\sigma$ defining the mapping between the CVH observables and the $\sl(2,\R)$ generators. On the other hand, a special point would be the critical values of $v$ and $b$ at the bounce, that is $\ka v=2\lambda\phi$ and $\cos^2\lambda b=\sin^2\lambda b=\f12$ for the choice of parameter $\sigma \ka^2\lambda =1$. In that case, the bounce corresponds to the value $\cJ_{z}=\ka^{-1}\pi_{\phi}/2$, which would correspond to $m=\f12$ at the quantum level.}
This means that, at the quantum level, the singularity is approached in the regime where the magnetic number $m \rightarrow - \infty$, while the classical limit (for a large volume universe) corresponds to $m \rightarrow + \infty$.

\medskip

At the end of the day, the main point that we would like to underline is that the $\SL(2,\R)$ symmetry is realized on the full phase space, including both gravitational and matter degrees of freedom. As a consequence, the two sectors, geometry and matter, are quantized together at once and are not decoupled as in the polymer quantization scheme.
On the one hand, the group quantization implements a $\SL(2,\R)$ action on quantum states of gravity plus matter, it allows for a straightforward quantization of the Hamiltonian constraint including the matter term with the inverse volume factor, but we do not have anymore direct access to a volume operator.
On the other hand, the polymer quantization realizes the $\SL(2,\R)$ symmetry only on the gravitation sector, uses the volume operator as its main building block but then faces the issue of quantizing the inverse volume factor with the ensuing necessary regularization, unavoidable quantization ambiguities and resulting possible anomalies.


\subsection{Imposing the dynamics and the choice of constraints}
\label{nosingularity}

In order to discuss the fate of the would-be singularity in the group quantization scheme, we need to first select the physical states and the associated scalar product and then study their asymptotic behaviour when $m \rightarrow 0$. Since all kinematical states live in a continuous representation $V_s$ at fixed spin $s$, they all satisfy the Casimir relation
\be
\mathfrak{C}|\Phi\ra= -\kappa^{-2}\hat{\pi}^2_{\phi} |\Phi\ra = -(s^2+1/4)|\Phi\ra
\ee
Now, when looking for the subset of physical state, an ambiguity shows up in the choice of the constraint to be imposed. Two choices appear to be equally justified.
\begin{itemize}
\item In order to impose the constraint associated to the deparametrized dynamics, one has to classically fix the gauge and work with the lapse function $N = V$. This leads to the classical constraint $H[V] = \kappa^{-2} \pi^2_{\phi} - \cC^2 \simeq0 $ which is equivalent classically to $\pi^2_{\phi} = \kappa^2 \cC^2$. This strategy is the one commonly adopted in the literature in quantum cosmology. At the quantum level, since $\hat{\pi}_{\phi}$ and $\hat{\cC}$ are well defined self adjoint operators, one can use the constraint 
\be
\kappa^{-2}\hat{\pi}^2_{\phi} |\Phi\ra =   \hat{\cC}^2 |\Phi\ra =  (s^2+1/4)|\Phi\ra
\ee 
to select the physical states and this is usually the strategy adopted. This is based on a classical equivalence.
\item However, one can argue that the operators $\hat{\cH}\hat{V}$ (or $\hat{V}\hat{\cH}$) are not self adjoint. From this point of view, one should work instead with the symmetrized version of this operator $\hat{\cH}\hat{V}+\hat{V}\hat{\cH}$ which is self adjoint. Using this constraint together with the CVH quantum bracket, one obtains then
\be
\Bigg{[}\f1{\ka^2}\big{(}\hcV\hcH+\hcH\hcV\big{)}+\hcC^2\Bigg{]}\,|\Phi\ra
=\Bigg{[}\f2{\ka^2}\hcV\hcH-i\hcC+\hcC^2\Bigg{]}\,|\Phi\ra
\,,
\ee
Imposing quantum mechanically that  $\hcH\,|\Phi\ra=0$, one obtains
\be
\forall |\Phi\ra\in V^s\,,\qquad
\hcH\,|\Phi\ra=0
\quad\Rightarrow\quad
\hcC(\hcC-i)|\Phi\ra =(s^2+1/4)|\Phi\ra
\ee
This last constraint is almost like the equation for the  deparametrized  dynamics, classically $\cC^2=\ka^{-2}{\pi}_{\phi}^2$. The subtlety is the quantum correction $-i\hcC$ with complex factor due to the operator ordering in the $\sl(2,\R)$ Lie alegbra. This means that physical states $\hcH\,|\Phi\ra=0$ lead to eigenvectors of the dilatation generator $\hcC$ as expected since $\cC$ is a classical constant of motion, but with a slight complex shift. This feature of working with states with complex eigenvalues of the $\sl(2,\R)$ generators might be awkward at first from a physical perspective  since the  $\sl(2,\R)$ generators $\hcK_{x,y}$ and $\cJ_{z}$ are self-adjoint in the unitary representation carried by $V^s$, but it is actually clear from a mathematical point of view. As explained for example in \cite{lindblad} (and references therein), the $\sl(2,\R)$ generators are self-adjoint operators on the space of rapidly decreasing states and their generalized eigenvectors live in the dual space, that is slowly increasing states, and not in the Hilbert space\footnotemark.
\footnotetext{
Working in the basis $|s,m\ra$ diagonalizing the generator of compact rotations $\hcJ_{z}$, the Hilbert space $V^s$ is the space of $L^2$ states:
\be
V^s=\Big\{|\Phi\ra=\sum_{m}\phi_{m}\,|s,m\ra;\,\, \sum_{m\in\Z}|\phi_{m}|^2<\infty\Big\}\,.
\nn
\ee
The space of rapidly decreasing states is defined as:
\be
\cD=\Big\{|\Phi\ra=\sum_{m}\phi_{m}\,|s,m\ra;\,\, \forall n\in\N\,,\,\,\lim_{|m|\rightarrow\infty}m^n\phi_{m}=0\Big\}
\,,
\nn
\ee
while its dual space is the space of slowly increasing states defined as:
\be
\cD'=\Big\{|\Phi\ra=\sum_{m}\phi_{m}\,|s,m\ra;\,\, \exists N\in\N\,,\,\,\lim_{|m|\rightarrow\infty}\f1{m^{N}}\phi_{m}=0\Big\}
\,.
\nn
\ee
The $\sl(2,\R)$ generators, $\hcJ_{z}$, $\hcK_{x}$ and $\hcK_{y}$, are well-defined and continuous on $\cD$ and leave $\cD$ invariant. Their action is extended to $\cD'$ by continuity since $\cD$ is dense on $\cD'$. Then the nuclear spectral theorem guarantees that there exists a complete set of generalized eigenvectors in $\cD'$.
}In that dual space, the generators, $\hcK_{x}$ and $\hcK_{y}$ actually admit eigenvectors for all complex eigenvalues\footnotemark. Nevertheless their spectrum is the real line $\R$ and eigenstates with real eigenvalues form a overcomplete basis of the Hilbert space providing a decomposition of the identity.
\footnotetext{
A simple example of similar behavior is the dilatation operator $-ix\pp_{x}-\f i2$ acting on smooth functions over the positive real line.}
\end{itemize}

%

%

%

%

Since both constraints are well defined, the difference is ultimately related on how they have been build. The first one is a self adjoint version of the constraint which is equivalent to the classical $\cH[V]\simeq 0$ constraint, while the second is a self adjoint operator consistent with the self adjoint quantum constraint $\hat{\cH} |\Phi\ra =0$. It follows from this discussion that both choice can be justified, resulting in an ambiguity in the constraint to impose in order to select the physical states. 

\medskip

Putting aside choosing one equation or the other to define the physical states of the theory, we would like to understand how to solve them and what are the generic properties of the resulting states. In general, the goal is to identify the eigenvectors of the operator $\hcC=\hcK_{y}$. A general method to compute the eigenvectors of the  $\sl(2,\R)$ generators using the Laplace transform is presented in \cite{lindblad}. Here it will be enough to understand the initial condition problems and the approximate asymptotic behavior in order to discuss boundary conditions at infinity and the resolution of the singularity.

\subsection{Solving the eigen-problem of the dilatation operator}
The goal is to solve in the $\SL(2,\R)$ representation of label $s$ the eigenvalue equation:
\be
\hcC  |\Phi\ra= \gamma |\Phi\ra \,,
\ee
with $\gamma=\pm s$ or $\gamma=\pm s+\f i2$. Decomposing the state in the $m$-basis, $|\Phi\ra=\sum_{m}\phi_{m}|s,m\ra$, and writing explicitly the action of the operator $\hcC=\hcK_{y}=(\hcK_{+}-\hcK_{-})/(2i)$, the eigenvalue problem is equivalent  to a second-order recursion relation on the coefficients $\phi_{m}$:
\begin{align}
\label{diffeq1}
 \forall m \in \mathbb{Z}\,,\qquad
 \beta_{m+1}\phi_{m+1} - \beta_{m} \phi_{m-1}  = -2i\gamma\phi_m \,,
\end{align}
where the coefficients $\beta_{m\pm1}$ are given explicitly by
\begin{align}
\beta_{m}  =  \sqrt{m (m-1)  - j(j+1)}  =  \sqrt{m (m-1)  +s^2+\f14}
\,.
\end{align}
The problem with the recursion on $\Z$ is that there is no special point where to define initial conditions but at infinity $\pm\infty$: we need to understand which asymptotic boundary conditions we can require on the $\phi_{m}$.
It is nevertheless possible to trade those boundary conditions for actual initial conditions, due to a special symmetry satisfied by the recursion coefficients:
\be
\beta_{m}  =  \sqrt{(m+j)(m-j-1)} 
\quad\Rightarrow\quad
\beta_{m}=\beta_{1-m}
\quad\Rightarrow\quad
\left|
\begin{array}{lcl}
\beta_{1}&=&\beta_{0}\\
\beta_{2}&=&\beta_{-1}\\
\beta_{3}&=&\beta_{-2}\\
\dots &&
\end{array}\right.
\ee 
This allows to map the evolution of the coefficients $\phi_{m}$ from $m\rightarrow -\infty$ to $m\rightarrow +\infty$ with boundary conditions at infinity to a initial condition recursion relation starting at $m=0$. We define the new sequences:
\be
\psi^\pm_{m}=\phi_{m}\pm\phi_{-m}
\,.
\ee
We combine the recursion relation for $\phi_{m}$ and its symmetric version for $\phi_{-m}$ into recursion relations for $\psi^\pm_{m}$, for $m\ge 1$:
\be
\left|
\begin{array}{lcl}
\beta_{m+1}\phi_{m+1} - \beta_{m} \phi_{m-1}  &=& -2i\gamma\phi_m 
\\
\beta_{m}\phi_{-m+1} - \beta_{m+1} \phi_{-m-1}  &=& -2i\gamma\phi_{-m} 
\end{array}\right.
\,\,\Rightarrow\,\,
\left|
\begin{array}{lcl}
\beta_{m+1}\psi^-_{m+1} - \beta_{m} \psi^-_{m-1}  &=& -2i\gamma\psi^+_m 
\\
\beta_{m+1}\psi^+_{m+1} - \beta_{m} \psi^+_{m-1}  &=& -2i\gamma\psi^-_{m} 
\end{array}\right.
\label{diffeq2}
\ee
We complete those equations with  the equations at $m=0$:
\be
\beta_{1}\phi_{1}-\beta_{0}\phi_{-1}=-2i\gamma\phi_{0}
\quad\Rightarrow\quad
\beta_{1}\psi^-_{1}=-i\gamma\psi^+_{0}
\,,
\ee
\be
\psi^-_{0}=\phi_{0}-\phi_{0}=0
\quad\Rightarrow\quad
\beta_{2}\psi^-_{2}=-2i\gamma\psi^+_1
\,. 
\ee
We realize that we have two decoupled sequences resulting from the recursion relations. On the one hand, $\psi^+_{0}$ determines $\psi^-_{1}$ and together they combine to give $\psi^+_{2}$, which in turn determines $\psi^-_{3}$, which determines $\psi^+_{4}$ and so on. On the other hand, $\psi^+_{1}$ determines $\psi^-_{2}$ and together they combine to give $\psi^+_{3}$, which in turn determines $\psi^-_{4}$ and so on.

We have turned the second order difference equation \eqref{diffeq1} on $\Z$ into two second order difference equations \eqref{diffeq2} on $\N$, each with half the normal initial conditions, thus with the correct expected number of initial conditions.
We can thus fix $\psi^+_{0}$ and $\psi^+_{1}$ and determine the whole sequences $\psi^\pm_{m}$.

\subsection{Asymptotic behaviour and singularity resolution}
\begin{figure}
\centering

\includegraphics[width=60mm]{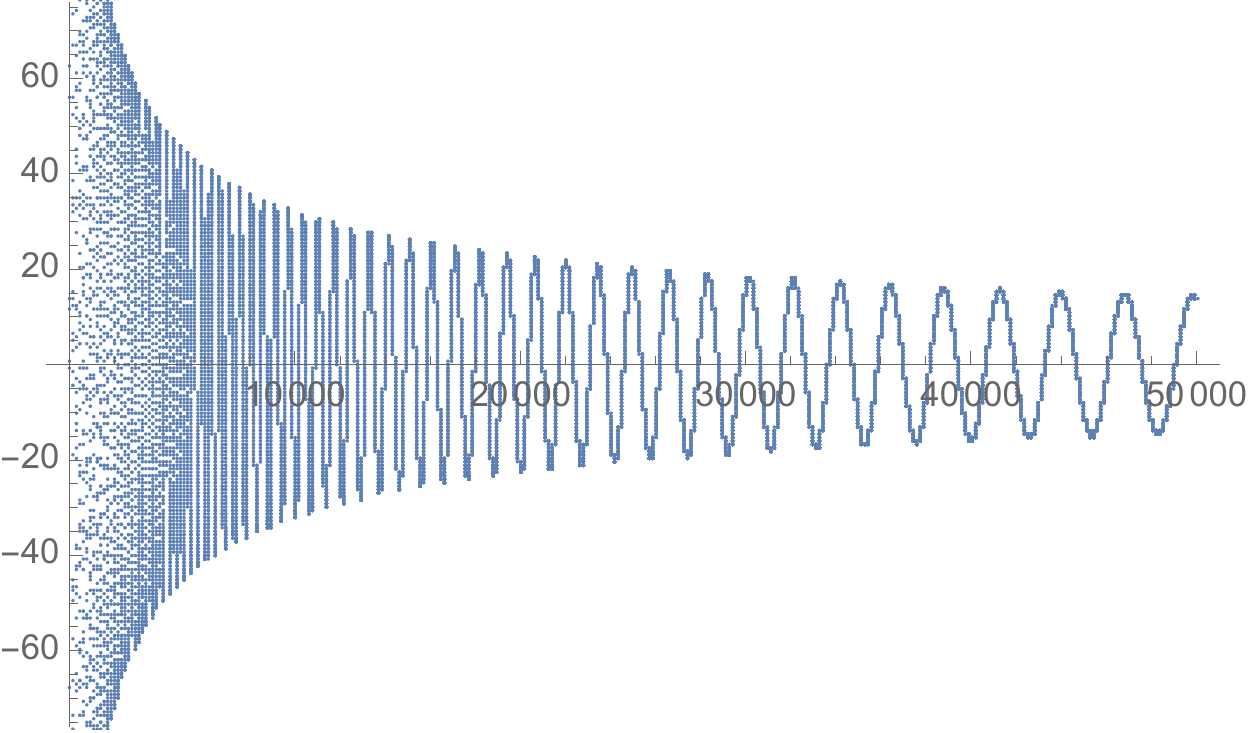}
\hspace*{10mm}
\includegraphics[width=60mm]{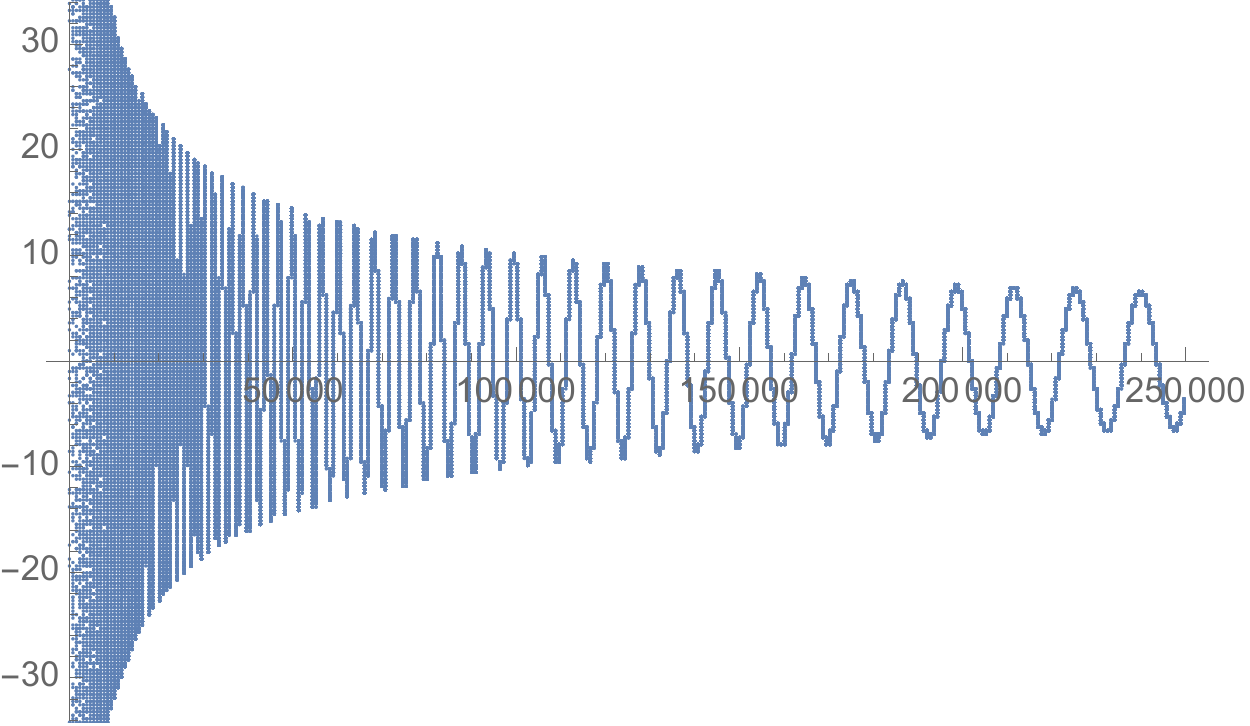}

\caption{Plots of the sequence $\vphi_{m}$ defined as the alternating sequence $\psi_{2m}^+,\psi^-_{2m+1}$ with initial condition $\psi^+_{0}=1$ and parameter $s=100$: we plot the even elements of the sequence $\phi_{2m}=\psi_{2m}^+$ for $m$ from 0 to 50000 on the left hand side, and for $m$ from 0 to 250000. We clearly see that the amplitude decreases and oscillates. The oscillations are scale-invariant (i.e. don't depend on the range of $m$) which indicates a log-behavior of the phase.}
\label{fig:recursion}
\end{figure}

\begin{figure}
\centering

\begin{subfigure}[t]{.40\linewidth}
\centering
\includegraphics[width=45mm]{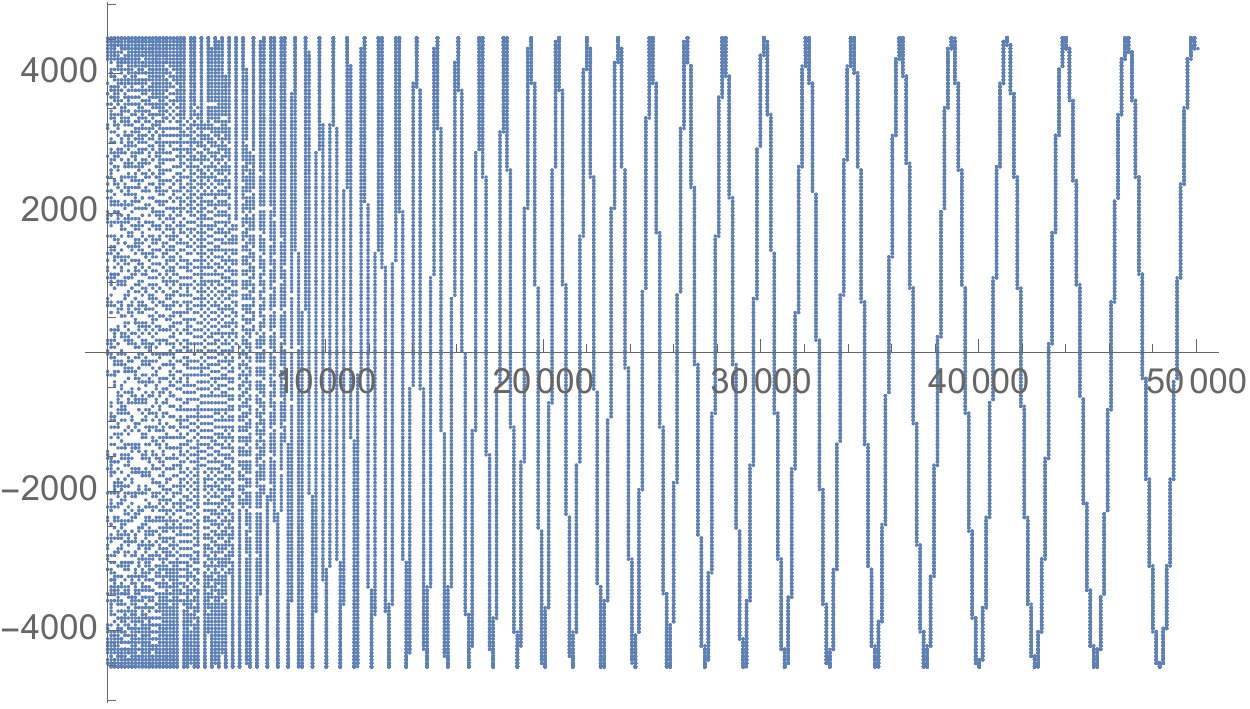}
\caption{Plot of $\vphi_{n}\sqrt{n}$ for even values $n=2m$ with $m$ ranging from 1 to 50000.}
\end{subfigure}
\hspace{5mm}
\begin{subfigure}[t]{.40\linewidth}
\centering
\includegraphics[width=45mm]{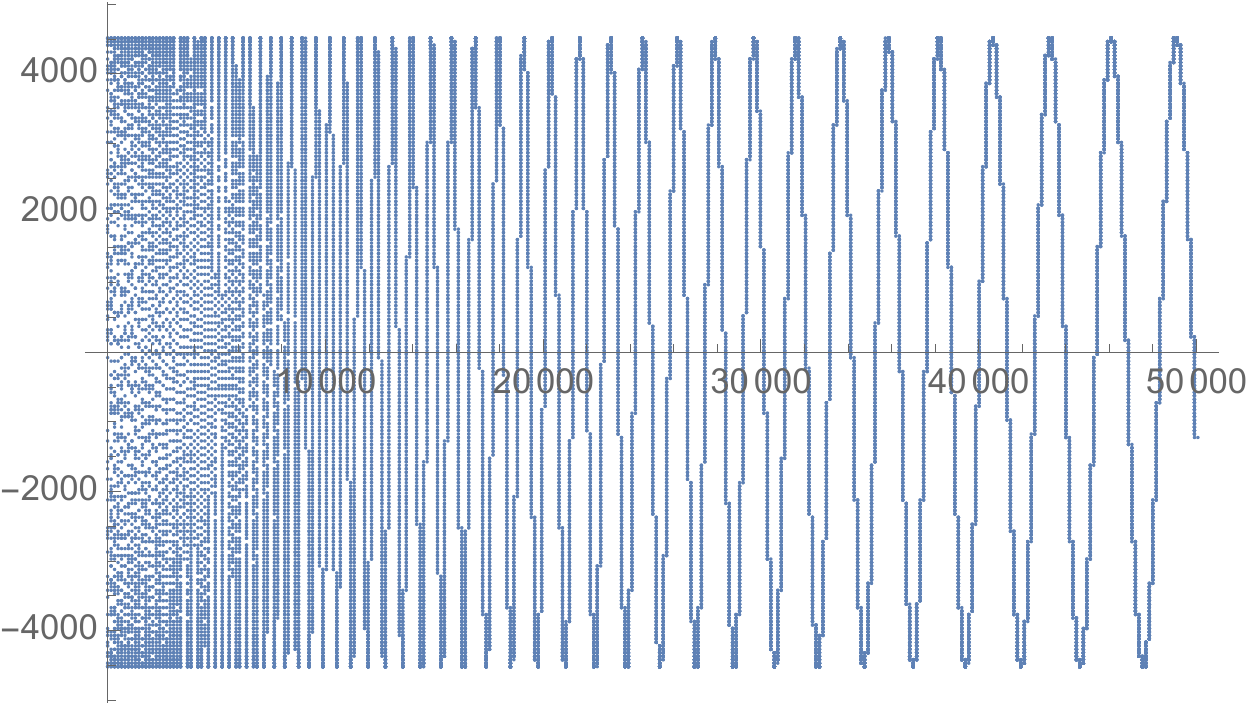}
\caption{Plot of $-i\,\vphi_{n}\sqrt{n}$ for odd values $n=(2m-1)$ with  $m$ ranging from 1 to 50000.}
\end{subfigure}
\\
\begin{subfigure}[t]{.40\linewidth}
\centering
\includegraphics[width=45mm]{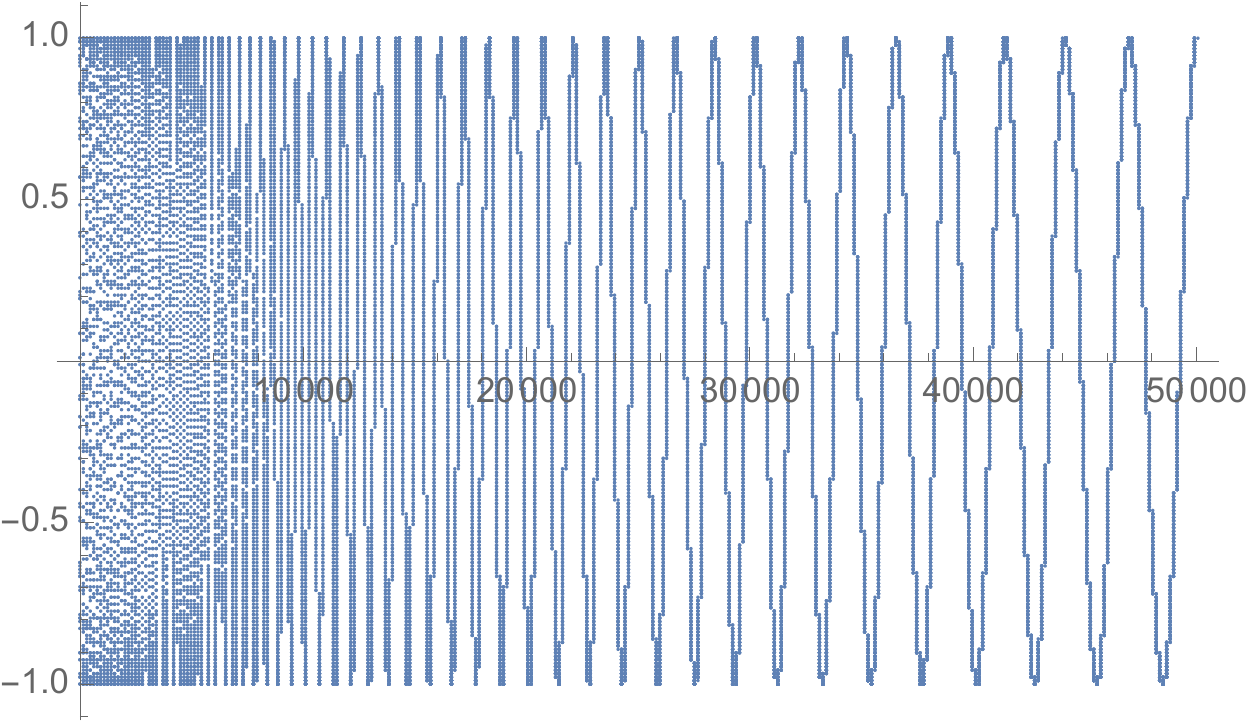}
\caption{Plot of $\sin[s\ln(2m)]$ as a fit of the oscillations of  $\vphi_{2m}\sqrt{2m}$.}
\end{subfigure}
\hspace{5mm}
\begin{subfigure}[t]{.40\linewidth}
\centering
\includegraphics[width=45mm]{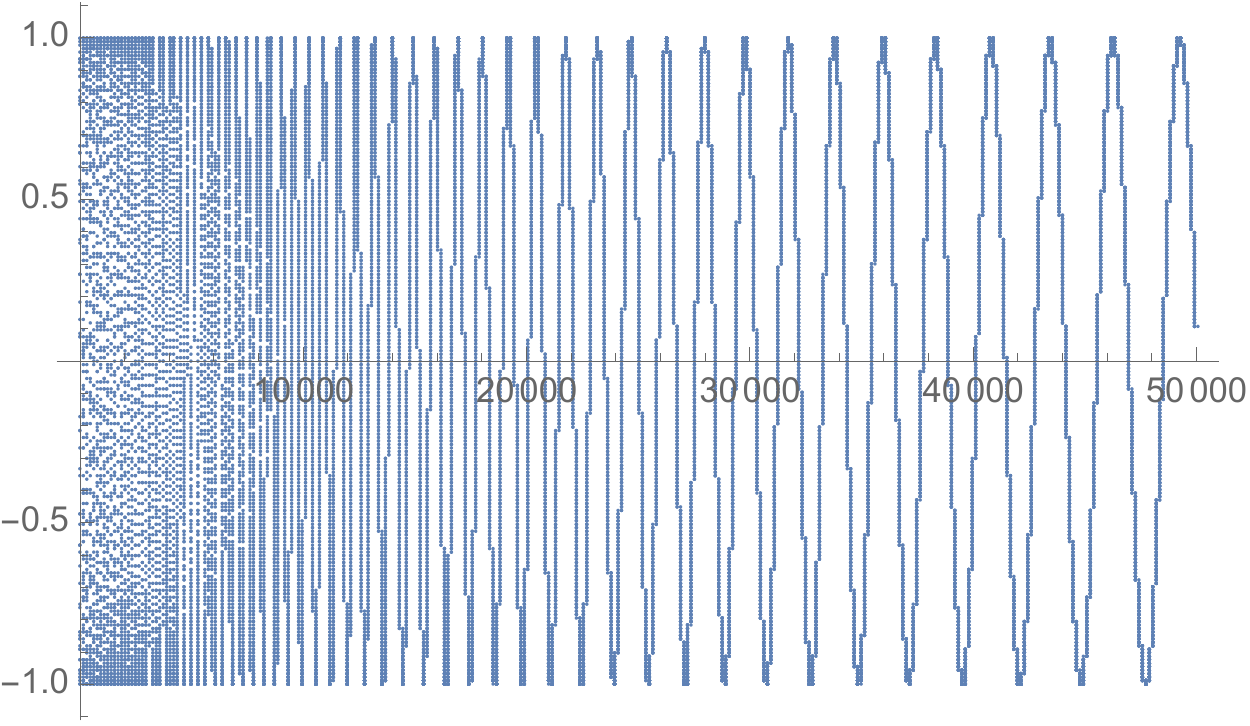}
\caption{Plot of $\cos[s\ln(2m-1)]$ as a fit of the oscillations of  $\vphi_{2m-1}\sqrt{2m-1}$.}
\end{subfigure}

\caption{Plots of the sequence $\vphi_{m}$ for $s=100$ rescaled by a factor $\sqrt{m}$.}
\label{fig:recursionfit}
\end{figure}

Now let us turn to the asymptotic behavior at $m\rightarrow +\infty$ of a sequence defined on $\N$ by the second order recursion relation:
\be
\beta_{m+1}\vphi_{m+1}=-2i\gamma \vphi_{m}+\beta_{m}\vphi_{m-1}
\,,
\ee
where the sequence $\vphi_{m}$ describes for instance the two alternating sequences of $\psi^{\pm}_{m}$.
For large $m$, the coefficient have a simple expansion:
\be
\beta_{m}\sim m-\f12 + {\cal O}\left(\f1m\right)
\,,
\ee
so the above recursion relation can be approximated by:
\be
\vphi_{m+1}\sim-2i\f\gamma m \vphi_{m}+(1-\f1m)\vphi_{m-1}
\,.
\ee
Assuming that the eigenvalue $\gamma$ is real, the two terms have different roles. The main contribution to $\vphi_{m+1}$ is $\vphi_{m-1}$ pondered by the decreasing factor $(1-\f1m)$. If this was the only contribution, the sequence $\vphi_{m}$ would straightforwardly converge to 0. The second contribution is a slight shift in the orthogonal  direction in the complex plane to  $\vphi_{m}$. This creates a slight rotation of the sequence $\vphi_{m}$. The decrease of the modulus $|\vphi_{m}|$ and the rotation of the phase $\textrm{Arg}(\vphi_{m})$ are combined in this recursion relation.
If we try a simple ansatz:
\be
\label{gggg}
\vphi_{m}\sim m^{\sigma+i\tau}
=
m^{\sigma}\,e^{i\tau\,\ln m}
\,,
\ee
with $\sigma$ and $\tau$ both real and respectively controlling the modulus and phase, we get up to $m^{-2}$ terms:
\be
\label{asymptfittt}
-2i\f\gamma m + \left(1-\f1m\right)\left(1-\f2m\right)^{\sigma+i\tau}\sim1
\quad\Rightarrow\quad
\left|
\begin{array}{lcl}
\sigma=-\f12 \\
\tau=-\gamma
\end{array}
\right.
\,.
\ee
The second independent asymptotic solution to the second order recursion relation is obtained by a sign flip:
\be
\vphi_{m}\sim (-1)^m\,m^{\tsigma+i\ttau}
\,,
\ee
which leads to:
\be
\tsigma=\sigma=-\f12
\,,\quad
\ttau=\gamma=-\tau
\,.
\ee
The specific linear combination of the two asymptotic solutions depend, a priori non-trivially, on the initial conditions at $m=0$.
So, for $\gamma\in\R$, this provides a prediction for the asymptotic behavior of the sequence $\vphi_{m}$ whose modulus is supposed converge to 0 as $1/\sqrt{m}$.
For instance, setting $\gamma=s=\ka^{-1}\pi_{\phi}$ as required by the Casimir condition, we look at the sequence $\psi_{0}^+$, $\psi^-_{1}$, $\psi^+_{2}$, \dots. Due to the initial recursion equation, $\beta_{1}\psi^-_{1}=-i\gamma\psi^+_{0}$, rescaling the initial condition  $\psi_{0}^+$ simply rescales the whole sequence. Therefor, we can set $\psi_{0}^+=1$, then $\psi_{1}^-$ and all the coefficients $\psi^-_{2n+1}$ for odd $m=2n+1\in2\N+1$ will be purely imaginary, while all the coefficients $\psi^+_{2n}$ will be purely real. 
We can check numerically the asymptotic behavior as:
\be
\psi_{2n}^+\sim \f1{\sqrt{2n}}\sin \big[s\ln (2n)\big]
\,,\qquad
\psi_{2n+1}^-\sim i\,\f1{\sqrt{2n+1}}\cos \big[s\ln (2n+1)\big]
\ee
This is illustrated on figures \ref{fig:recursion} and \ref{fig:recursionfit} for the numerical value $s=100$.
This means that all the $\psi^\pm_{m}$ converge to 0 as $m$ goes to $\infty$. In particular, this implies that the sequence of original coefficients of the quantum state $\phi_{m}$ behave as
\be
\label{singreso}
\lim_{m \rightarrow - \infty} \phi_{m}=\lim_{m \rightarrow + \infty} \left(\psi^{+}_{m} - \psi^{-}_{m}\right)  = 0
\ee
thereby avoiding an accumulation of the quantum state at the zero volume state and avoiding the singularity. This provides thus a singularity resolution mechanism in this group quantization scheme. Once again, notice that the behavior of the wave function near the would-be singularity, i.e (\ref{singreso}) is \textit{derived} in this quantization, and does not descend from ad hoc boundary condition on the wave function.

Let us conclude this section with an important comment on the choice of Hamiltonian constraints for the physical states. Indeed, here we choose to solve the eigenvalue problem for $\gamma=s\in\R$. If we had chosen the shifted eigenvalues, $\gamma=s+\f i2 \in \mathbb{C}$, the argument above would be slightly modified, nonetheless leading to a completely different output. Indeed, the equation \eqref{asymptfittt} for the asymptotic ansatz would give a different result:
\be
\label{asymptfit}
i\gamma+\f12+\sigma+i\tau=0
\quad\Rightarrow\quad
\left|
\begin{array}{lcl}
\sigma=0 \\
\tau=-s
\end{array}
\right.
\,.
\ee
This would mean no decay\footnotemark{} at $m$ goes to $\infty$ and a purely oscillatory behavior for the state coefficients $\phi_{m}$.
\footnotetext{The ansatz $m^\sigma$ for $\sigma$  nevertheless allows for a possible logarithmic decay or divergence. The precise fate of the asymptotic behavior remains to be studied thoroughly, most likely using the exact methods for second order recursion as described in \cite{lindblad}.} It is worth pointing out that, the dilatation operator being hermitian, the decomposition of the identity can be realized in terms of the eigenstates associated to real eigenvalues $\gamma \in \mathbb{R}$. However, while it is also possible to write a decomposition of the identity using another contour in the complex plane of eigenvalues $\gamma \in \mathbb{C}$, the integration has be performed with a different measure, and thus a modified scalar product. For this reason, despite the apparent singularity for the second choice of constraint, no conclusion can be drawn at this level when $\gamma$ is shifted in the complex plane, $\gamma \in \mathbb{C}$, and further investigation is needed to conclude in that second case.
%

\subsection{Comparing the Polymer and the $\SL(2,\R)$ Quantizations}

Let us compare the present quantization scheme, preserving the full $\SL(2,\R)$ symmetry for the coupled system gravity plus matter, to the polymer quantization presented in the previous section preserving the $\SL(2,\R)$ symmetry only in the gravitational sector. They are a priori inequivalent quantization scheme since they do not quantize the same algebra of observables. One key observable is nevertheless common to the two schemes: the dilatation generator $\hcC$. We can thus compare how the two quantization procedure solve the (deparametrized) Hamiltonian constraint $\hcC=\ka^{-1}\pi_{\phi}$.

On the one hand, the polymer quantization proposes a Hilbert space spanned by states $|n\ra$ with $n\in\N$ diagonalizing the volume operator,
$\hv|n\ra=2\lambda n |n\ra$ and quantizes the constraint $\hcC|\Psi\ra=\ka^{-1}\pi_{\phi}|\Psi\ra$ as the following recursion relation, where we rescaled the recursion relation \eqref{eqdiff2} given earlier by a factor 2 to work with $n\in\N$ instead of $n\in2\N$ :
\be
|\Psi\ra=\sum_{n\in\N}\Psi_{n}|n\ra
\,,\qquad
-2i\f{\pi_{\phi}}\ka\Psi_{n}
=
\sqrt{n(n+1)}\Psi_{n+1}-\sqrt{n(n-1)}\Psi_{n-1}
\,.
\ee

On the other hand, the $\SL(2,\R)$ group quantization scheme proposes a Hilbert space spanned by states $|m\ra$ with $m\in\Z$ diagonalizing the $\sl(2,\R)$ generator $\hcJ_{z}|m\ra=m|m\ra$. This observable is not exactly the volume operator and contains a matter term involving an inverse volume factor (which plays a crucial near the zero volume singularity):
\be
\cJ_{z}=
\frac{v}{2 \lambda} 
-
\frac{\lambda  \pi^2_{\phi}}{2\ka^2 v \cos^2{(\lambda b)}}
\,.
\nn
\ee
This expression also involves a regularization factor $\cos^2{(\lambda b)}$ depending on the conjugate variable $b$. The Hamiltonian constraint equation $\hcC=\ka^{-1}\pi_{\phi}$ now leads to a similar recursion relation, given earlier in \eqref{diffeq1}:
\be
|\Phi\ra=\sum_{m\in\Z}\Phi_{m}|m\ra
\,,\quad
-2i\f{\pi_{\phi}}\ka\Phi_{m}
=
 \sqrt{m (m+1)  +s^2+\f14}\Phi_{m+1}
 -
  \sqrt{m (m-1)  +s^2+\f14}\Phi_{m-1}
  \,,
  \nn
\ee
with $s=\pi_{\phi}/\ka$.

The difference between the two recursion relation is clear. From a mathematical perspective, there are two differences: the $(s^2+1/4)$ correction under the square-root in the recursion coefficients and the range of the state labels, $n\in\N$ versus $m\in\Z$.
First the $(s^2+1/4)$ correction comes from using a different $\SL(2,\R)$-representation. The polymer quantization uses a null representation, with vanishing Casimir $\mathfrak{C}=0$, while the the group quantization uses a space-like representation with negative Casimir $\mathfrak{C}=-(s^2+1/4)$ with the representation label $s=\ka^{-1}\pi_{\phi}$.
Second this correction term is directly responsible for the difference of range. Indeed the null representation, initially also defined with $n\in\Z$, is actually reducible and splits into two irreducible representations with range $n\in\N^*$ and $n\in-\N^*$. We can also see this directly from the recursion relation above where positive values of $n$ naturally decouple from negative values. The difference of range $n\in\N$ versus $m\in\Z$, is not problematic because $n$ and $m$ are not the eigenvalues of the same observables. Indeed the zero volume corresponding to $n=0$ in the polymer Hilbert space  is sent to $m=-\infty$ in the group quantization scheme. This is actually an elegant feature of the $\SL(2,\R)$ group quantization that the issue of the zero volume singularity is sent to infinity. Actually the whole range of negative values $m<0$ can be understood as probing subPlanckian volume with $v$ between 0 and the regularization scale $\lambda$.

From a physical perspective, the $(s^2+1/4)$ correction, with $s=\pi_{\phi}/\ka$, can be considered as an extra layer of quantum gravity corrections. The matter and geometry are now intertwiners in the definition of the observables and the dynamics, and the non-linearity in $\pi_{\phi}$ of the quantum Hamiltonian constraint can be interpreted as taking into account the non-perturbative feedback of matter on the (quantization of the) geometry.

At the end of the day, let us underline that the $\SL(2,\R)$ quantization scheme allows to recover all the main features of loop quantum cosmology (LQC), quantization of a volume-like observable, singularity resolution and cosmological bounce driven by quantum gravity effects, without using the polymer quantization.

\section{Discussion}
\label{D}

We have shown that the simplest model of classical cosmology, invariant under time repara\-metrization, enjoys a residual conformal structure. We have emphasized that such hidden $\sl(2,\mathbb{R})$ structure can be used as a new powerful criteria to restrict the quantization ambiguity in quantum cosmology. This structure being an in-built structure of the classical system, and since it fully encodes the cosmological dynamics, it appears as a crucial ingredient to preserve in the quantum theory. This new criteria can be applied to any quantization scheme, being the standard Wheeler-De Witt quantization based on the Schrodinger representation, or the loop quantization based on the polymer quantization. We believe that it could also provides an interesting criteria in the path integral approach to quantum cosmology.

In this work, we have focused our attention of the loop regularized $\sl(2,\mathbb{R})$ invariant cosmological system. We have followed two different paths to work out a consistent quantization of the effective phase space: a standard polymer (or loop) quantization, and a group quantization. As such, this work provides a further generalization of our previous study presented in \cite{BenAchour:2017qpb}, where a group quantization was only briefly sketched. The construction developed here is summarized in a companion paper \cite{BenAchour:2018jwq}. Let us now emphasize the main results obtained in the present generalization. 

At the level of the polymer quantization procedure, this hidden conformal structure constrains the regularization of the extrinsic curvature inherent to the polymer quantization, leading to a new LQC model with several improvements. Scale transformation of the three-dimensional hypersurface are now generated by a unitary quantum operator $\widehat{\cC}$, which furthermore coincides at the classical, effective and quantum level with the generator of the deparametrized dynamics. As a consequence, the minimal length introduced in the quantization remains a universal scale, contrary to the standard construction\footnote{This can be viewed as a resolution of the Immirzi ambiguity in this simple symmetry reduced loop quantized model}. The present model provides therefore an explicit example of how to reconcile the existence of a universal minimal scale in quantum cosmology with a 3d scale invariance at the quantum level. It can be understood as a non-linear mapping of FLRW cosmology, allowing to resolve the big bang singularity into a regular quantum cosmology. 

The preservation of the conformal structure requires new holonomy-corrections (i.e. factors depending on the extrinsic curvature $b$) to the (inverse) volume term, which do not seem to coincide with the standard triad corrections already investigated in LQC and obtained through the Thiemann's trick \cite{Bojowald:2001vw}. However, we point that in standard LQC, one does not exactly proceed to a loop quantization since the covariant fluxes operators are not implemented. Yet, the covariant version of the fluxes variable contains also a non-trivial dependency of the extrinsic curvature, and thus some holonomy. While this is usually ignored in LQC, we wonder whether our new corrections could originate from this property of the fluxes variables in such symmetry-reduced cosmological model. This remains to be checked. To our knowledge, this point has not yet been investigated in LQC. Finally, let us point that the holonomy-corrections to the inverse volume term in our new hamiltonian have some similarities with the alternative regularization introduced in \cite{Campiglia:2016fzp} in the context of gravitational collapse. We believe that the new model presented in this work provides a more geometrical construction than its standard counterpart, since its quantization is rendered unambiguous due to the central role of the  $\sl(2,\R)$ algebra. Moreover, as shown recently in \cite{Bodendorfer:2018csn}, such conformal structure can be used to develop efficient coarse-graining method based on properties of $\SU(1,1)$ coherent states. Additionally, the present construction could also provides a simplified framework to construct a Wick rotation aiming at defining the quantum theory in term of the self dual variables, as proposed initially in \cite{Thiemann:1995ug}. See \cite{Achour:2014rja, Wilson-Ewing:2015lia, Wilson-Ewing:2015xaq, Frodden:2012dq, Achour:2014eqa, Achour:2015xga, BenAchour:2016mnn} for more details on different attempts to develop this program in LQC and black hole state counting.

We stress finally that the new symmetry criteria based on the conformal structure is not tied to the high degrees of symmetry of flat homogeneous and isotropic cosmology, but holds also for more general models. As shown in Appendix-\ref{BianchiI}, this structure can also be used to constraint the polymer quantization of Bianchi I model \cite{Wilson-Ewing:2017vju}. This is especially interesting towards discussing the growth of anisotropies in such lattice like structure \cite{Pioline:2002qz}. A more general investigation of this symmetry at the level of General Relativity could reveal an interesting hidden structure with far reaching consequences. We plan to study this crucial point in the future.

We have also presented a complementary group quantization of the same loop regularized system in Section-\ref{C}. This alternative scheme, while not relying on the polymer hypothesis, allows to obtain a well defined quantum cosmology with a singularity resolution derived from the scalar constraint. The structure of the theory appears to be more geometrical in this scheme, as the $\sl(2,\mathbb{R})$ structure can be fully realized on both the gravitational and matter sector at once, contrary to the loop quantization where the conformal structure is realized only in the gravitational sector. This crucial difference is best illustrated by the fact that the operator with discrete spectrum is not the same in the two schemes. In the loop quantization, this operator corresponds to the physical volume of the universe, leading to a lattice like structure of the geometry. On the contrary, in the group quantization scheme, the rotation generator of the $\sl(2,\mathbb{R})$ algebra is identified to a mixed combination of gravitational and matter degrees of freedom, and the volume operator is no more the fundamental discrete quantity. In this sense, the gravitational and matter degrees of freedom are less easily disentangled in the group quantization picture, leading to a more covariant picture of the quantum universe. 

From a more general perspective, our model provides a lattice-like quantum cosmology with an $\sl(2,\R)$ invariance. In principle, the presence of this conformal symmetry at the quantum level could allow to recast this lattice-like quantum cosmology as a generally covariant 1D conformal quantum field theory \cite{CFT}. It implies that one could derive the exact form of the correlators at all order solely based on its conformal symmetry and thus bootstrap this quantum cosmological system \cite{Kitaev:2017hnr, McElgin:2015eho}. Additionally, it would be also interesting to investigate whether the existence of this conformal symmetry at the quantum level allows a mapping towards other quantum mechanical systems of interests, among which the well known conformal quantum mechanics (CQM) developed by de Alfaro, Fubini and Furland \cite{deAlfaro:1976vlx, Andrzejewski:2011ya, Andrzejewski:2015jya}, as well as the large N limit of the SYK model \cite{Rosenhaus:2018dtp, Sarosi:2017ykf, Maldacena:2016hyu}. Both of these models enjoy a conformal symmetry which fully determine the $n$-points correlators  and thus allow to fully solve the quantum theory using the conformal bootstrap \cite{Mertens:2017mtv}. Whether our $\sl(2,\R)$ invariant lattice-like quantum cosmology model could emerge in the large N limit of a more fundamental quantum mechanical matrix or tensor models such as \cite{Stern:2018wud, Ho:2004qp, Axenides:2013iwa, Pinzul:2017wch} remains to be investigated. This could also be useful in order to build a consistent holographic description of this quantum cosmology \cite{Okazaki:2015lpa}.

As a last remark, we point that the recent construction of GFT cosmological condensate based on the Gross-Piteavskii approximation could find an alternative realization within our new model \cite{Baytas:2018qty,  Gielen:2013naa, Oriti:2016ueo, Oriti:2016acw}. Indeed, the $\SL(2,\mathbb{R})$ structure of our quantum cosmology suggests interesting mapping with the so-called conformally invariant quintic non-linear Schrodinger (NLS) equation. Such approach has already been initiated in \cite{Lidsey:2018byv}. These NLS equations describe low-dimensional condensates with non standard properties which appear to be relevant for quantum gravity models. We plan to investigate these different interconnections opened by our new construction in the future.

\section*{Acknowledgments}
This work was supported by the National Science Foundation of China, Grant No.11475023 and No.11875006 (J. BA), as well as by the China Postdoctoral Science Foundation  with Grant No. 212400209 (J. BA).

\appendix

\section{Classical conformal structure of the Bianchi I model}
\label{BianchiI}

In this appendix, we show that the conformal $\sl(2,\mathbb{R})$ structure is not an accidental property of the higher degrees of symmetry of the homogeneous and isotropic mini-superspace. On the contrary, it holds for a larger class of cosmological systems of interest. Here, we consider an anisotropic cosmological background given by the Bianchi I model, minimally coupled to a massless scalar field. The loop quantization of this model was performed in \cite{Ashtekar:2009vc, MartinBenito:2011fv, Singh:2011gp, MartinBenito:2009qu, Singh:2012zzj, Fujio:2013xwa}.  

The Bianchi I geometry admits the topology $\Sigma \times \mathbb{R}$, where the hypersurface $\Sigma$ can have compact or non compact directions. Working with the standard topology $\Sigma = \mathbb{R}^3$, the metric takes then the standard form
\be
ds^2 = - N(t)^2dt^2 + \sum_{i=1}^3 a^2_i(t)  (dx^i)^2
\,.
\ee
Integrating over a fiducial 3D cubic cell of edge length $\ell_{\circ}$ in the $x_{i}$'s  in order to avoid divergence due to the non-compact spatial directions, its 3D volume is given by:
\be
v = \int_{\Sigma_{\circ}} \sqrt{\gamma} = V_\circ \,a_1 a_2 a_3
\,,
\quad\textrm{with}\quad
V_{\circ}=\ell_{\circ}^3
\ee
and the reduced Einstein-Hilbert action for the coupled homogeneous gravity-scalar system is obtained by integrating the scalar curvature:
\be
\cS=V_{\circ} \int dt\, \Bigg{[}
\f{a_{1}a_{2}a_{3}}{2N}\dot{\phi}^2
-\f1{8\pi G}\f1N\sum_{i=1}^3a_{i}\dot{a}_{j}\dot{a}_{k}
\Bigg{]}\,,
\ee
where we are using the notation  $j,k$ to denote the two other indices in $\{1,2,3\}$ other than $i$ (i.e. $(j,k)=(2,3)$ for $i=1$). Strictly speaking, we should write $j<k$, $j\ne i$, $k \ne i$.
When the three scale factors are equal, $a_{i}=a(t)$ for all $i=1,2,3$, we recover the action for the FLRW cosmological model that we have used in the body of the paper.

It is convenient to work with variables $\alpha_i = \ln a_i$, $a_{i}=e^{\alpha_{i}}$, writing as before $\ka=\sqrt{12\pi G}$,
\be
\cS=V_{\circ} \int dt\, \Bigg{[}
\f{e^{\alpha}}{2N}\dot{\phi}^2
-\f{3e^{\alpha}}{2\ka^2\,N}\sum_{j<k}\dot{\alpha}_{j}\dot{\alpha}_{k}
\Bigg{]}
\,.
\ee
We compute their conjugate momenta,
\be
\Pi_{i}
=
-\f{3V_{\circ}e^\alpha}{2\ka^2\,N}\sum_{j\ne i}\dot{\alpha}_{j}
=
-\f{3v}{2\ka^2\,N}\sum_{j\ne i}\dot{\alpha}_{j}
\,,
\qquad
\{\alpha_{i},\Pi_{j}\}=\delta_{ij}
\,.
\ee
The Hamiltonian is then a constraint, enforced by the lapse $N$ as Lagrange multiplier:
\be
H[N]
=
\f{N}{2v}\Bigg{[}
\pi_{\phi}^2
-\f{\ka^2}3\Big{[}
2(\Pi_{1}\Pi_{2}+\Pi_{2}\Pi_{3}+\Pi_{3}\Pi_{1})-(\Pi_{1}^2+\Pi_{2}^2+\Pi_{3}^2)
\Big{]}
\Bigg{]}\,.
\ee
Note that the inverse volume term in the kinetic gravitational term arises because of the use of the canonical variables $(\alpha_{i}, \Pi_{i})$.

Once again, we find that the Hamiltonian $H$ and the 3D volume $v$ generate a finite dimensional Lie algebra descending from the Poisson bracket structure. We compute the first Poisson bracket:
\be
\{ v , H[N] \} =  - N\f{\ka^2}3 \left( \Pi_1 + \Pi_2 + \Pi _3\right) \equiv N \ka^2 C\,,
\ee
which we take as the definition for the new observable $C$. It turns out that $C$ is the integrated extrinsic curvature\footnotemark{} and acts, as the dilatation generator on the Hamiltonian and the volume, as in the FLRW case,
\be
\{C,v\}=+v
\,,\qquad
\{C,H[N]\}=-H[N]
\,.
\ee
\footnotetext{
We can compute the extrinsic curvature of the space-like hypersurface, $K_{ii}=a_{i}\dot{a}_{i}/N$,  its  trace $K=N^{-1}\sum_{i}\dot{a}_{i}/a_{i}$ and its integration over the 3D fiducial cell:
\be
\int_{V_{\circ}} a_{1}a_{2}a_{3}\,K
=
\f{V_{\circ}e^\alpha}{N}\sum_{i}\f{\dot{a}_{i}}{a_{i}}
=
-\f{\ka^2}3\sum_{i}\Pi_{i}
=
\ka^2\, C
\,,
\nn
\ee
which thus gives the dilatation generator up to the Planck area  factor $\ka^2$.
}
Hence, we obtain the same $\sl(2,\mathbb{R})$ symmetry algebra even in presence of anisotropies. One can proceed to the same identification between the $\sl(2,\R)$ and the CVH generators. The $\sl(2,\mathbb{R})$ Casimir is then again given by 
\be
\label{CazMat}
\mathfrak{C}_{\mathfrak{sl}_{2}}
= j^2_z - k^2_x - k^2_y 
= -\f2{\ka^2}vH-C^{2}
= -  \frac{\pi^2_{\phi}}{\ka^2} < 0
\ee
The technics developed for the simple massless scalar minimally coupled to isotropic and flat FLRW geometry can therefore be generalized to the Bianchi I cosmology. The next step is to generalized the matter couplings but we leave this for future work. However, we note that it is straightforward  to include a minimally coupled Maxwell field within the Bianchi I cosmology  without breaking the CVH algebra.


\end{document}